\documentclass[prl,twocolumn,showpacs,preprintnumbers,amsmath,amssymb]{revtex4}
\usepackage{graphicx} 

\usepackage[T1]{fontenc}
\usepackage[ansinew]{inputenc} 

\usepackage{comment}
\usepackage{overpic}
\usepackage{mathtools}

\usepackage{amsmath}

\usepackage{textcomp}
\usepackage{gensymb} 

\usepackage{lipsum,mathptmx,etoolbox}

\usepackage{calc}
\usepackage{graphicx}
\usepackage{amsmath}
\usepackage{amssymb}
\usepackage{gensymb}
\usepackage{epsfig}
\usepackage{amsthm}
\usepackage{bm}
\usepackage{color}
\usepackage[toc,page]{appendix}
\usepackage{ragged2e}

\usepackage{subfigure}

\usepackage{overpic}
\usepackage{float} 

\usepackage{xr}

\def\be{\begin{equation}}	
\def\ee{\end{equation}}
\def\arr{\begin{array}{rll}}
\def\ea{\end{array}}
\def\bea{\begin{eqnarray}}
\def\eea{\end{eqnarray}}

\begin{document}

\title{Strong Energy Dependent Transition Radiation in a Photonic Crystal}
\author{V.Gareyan$^{1}$ and Zh.Gevorkian$^{1,2}$}
\affiliation{$^{1}$ Alikhanyan National Laboratory, Alikhanian Brothers St. 2,  Yerevan 0036 Armenia\\
$^{2}$ Institute of Radiophysics and Electronics, Ashtarak-2 0203 Armenia}
\begin{abstract}
Radiation of a charged particle crossing an alternating stack of slabs in the optical region is considered. Both disordered and periodic stacks are investigated. It is shown that for special type of alternating disordered  and periodic stacks the radiation problem can be solved exactly for backward and forward Brewster observation angles. Strong $N^2$ dependence of radiation intensity on slab  number is re-established in special case of the disordered stack. This leads to strong directivity either on forward or on backward Brewster angles depending on the type of stack randomness. In certain type of periodic photonic crystal, a strong energy dependence $E^4$ for relativistic particles of the radiation intensity, observed at Brewster's angle is found. Further increment of particle energy leads to saturation. The band structure of the corresponding photonic crystal (PhC) has a behavior, analogous to the Dirac cones in graphene. We suggest this special type $1D$ photonic crystal for application as a detector of relativistic particles.

\end{abstract}

\maketitle

\textit{Introduction}- Recently the interest to the radiation of charged particle in layered media is resumed \cite{pnas2025,yang25,fur25,ding25}. This is caused by the possibility of fabrication of necessary structures at nanoscale nowadays. Radiation from a charged particle can be used for many purposes: including,  as a light source \cite{Liao2016,casal09,tak25,hof11,tay23,kang23,sei17} in those wavelength regions where other methods are ineffective, detection of relativistic charged particles \cite{AndWes,dol93,art75,alice18,det88}, in bio-medical applications \cite{clot22,pok2006} etc. In early works on transition radiation \cite{Garib,TerMik,Potylitsyn,Ginzburg}, for a recent review see \cite{Lin}, it was believed that to achieve directivity and high intensity one needs to use regular structures. However, they had certain disadvantages: since the observations were performed in the X-ray region, the maximal intensities were observed in the vicinity of the particle trajectory and the efficiency of these detectors was limited while measuring large Lorentz factors because of fast growth of the radiation formation zone length.

Nevertheless, attention to radiation in irregular structures \cite{Chen23,gev,gevexp} was also paid. The main peculiarity in these random systems is the emergence of the diffusive mechanism of radiation \cite{gev,gevexp} along with the ordinary transition radiation.
In \cite{Chen23} the final conclusion was to investigate the aperiodic structures as possessing "more degrees of freedom to tailor the particle-interface interaction" compared to periodic ones. The Brewster's angle was highlighted in \cite{pnas2025} as the one, which allows to achieve maximal intensity in observations because of the coherent interference of a particle radiation from each boundary trespassed. However, the other important quantity for the detector, the strong dependence on Lorentz factor of the trespassing particle, was not considered. In fact, no ultrarelativistic particle was considered. 

In periodical case \cite{Lin} the possibility of generation of Cherenkov like directed radiation in forward and backward directions was demonstrated. The radiation is concentrated inside the angles close to particle trajectory and is sensitive to its energy. The energy dependence of varying radiation angle is suggested to be implemented for detection of relativistic particles. However, the Brewster's angle was out of the scope of the article, main focus was on the observation directions, way closer to the particle trajectory. Besides, the efficiency in measuring the Lorentz factor was limited even more, compared to the traditional X-ray RTR detectors.

Here we suggest an alternative approach. Considering radiation from a charged particle  at fixed Brewster observation angles we get strong energy dependence of radiation intensity for special type $1D$ photonic crystals. It is well known that ordinary transition radiation and Cherenkov radiation in optical region slowly depend on particle energy. We find a certain conditions, which will allow to tune the effective region of Lorentz factors for ultrarelativistic particles to be measured.

\begin{figure}[H]
    \centering
    \includegraphics[width=0.9\linewidth]{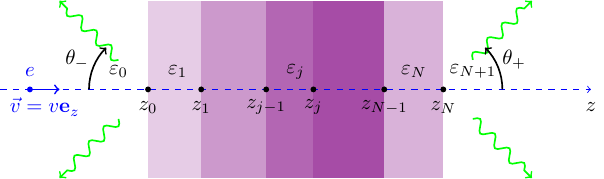}
    \caption{Illustration of the problem}
    \label{illustration}
\end{figure}

\textit{Theory}- We consider a stack on $N$ dielectric slabs with dielectric constants $\varepsilon_i, \forall i = \overline{1,N}$, separating the media with dielectric constants $\varepsilon_0$ from the left and $\varepsilon_{N+1}$ from the right, see Fig.\ref{illustration}. A particle with charge $e$ traverses the system normal to each slab and with constant velocity $v = c\beta$. We need to calculate the distribution of the radiated power.

This reduces to the following wave equation for the magnetic field:

\begin{gather}
    \left[ \frac{\omega^2}{c^2}\varepsilon(z, \omega) - q^2 +\partial_z^2 \right] {\bf H} (z, {\bf q}, \omega) = -\frac{4\pi i e}{c} ({\bf q} \times {\bf e}_z) e^{i\frac{\omega}{v}z}
    \label{waveeq}
\end{gather}

, where we performed the Fourier transformation, arriving to the component, corresponding to the transversal momentum ${\bf q}$ and the frequency $\omega$. Note, that the magnetic field lies everywhere in the plane, parallel to the slabs. This means, that the radiation in the system is always $p$-polarized. Because we consider a stratified medium, we can decompose the total magnetic field into a linear combination of forward and backward-propagating pseudo-photon fields \cite{Ginzburg} inside each medium $j=0,...,N+1$ with the respective amplitudes $A_j^+$ and $A_j^-$. It is also taken into account that outside the system ($j=0$ or $j=N+1$) only the following components survive:

\begin{gather}
    A_{N+1}^+ = - B_1/M_{11}, \nonumber\\ A_0^- = B_2 - \frac{M_{21}}{M_{11}} B_1,
    \label{mateq}
\end{gather}

where $\hat{M}$ is the transition matrix for the entire system and $\hat{B}$ is the current-indiced term, given in the Supplemental Material. 

Afterwards, the total field is decomposed into two components ${\bf H} = {\bf H}^e + {\bf H}^r$: the one, corresponding to the charge, which is proportional to $\exp(iz\omega/v)$, and the one, corresponding to the radiation, proportional to $\exp(\pm ik_z z)$. With this, 3 components arise in the Pointing vector ${\bf S} \sim {\bf E} \times {\bf H}$: composed from 1) both charge fields, which reduces to the Cherenkov radiation, once $v>c/\sqrt{\varepsilon}$ or in case of superluminal velocities in the medium 2) both radiation fields, which reduces to the transition radiation, and 3) the mixed term, composed from the charge and radiation fields. The latter consists of the components, oscillating in space through a so-called radiation formation zone lengths $z^{\pm}(\theta) = \pi/|\omega/v \mp k_z(\theta)|$, corresponding to the forward (+) and backward (-) propagating components. The latter vanishes only if the detector is placed far enough ($\gg z^{\pm}$) from the radiator.

Such a general statement of the problem as in Eq.
(\ref{mateq}) allows to consider various particular cases. For example, by solving the problem in case of 2 media, separated by a plane interface ($N=0$),  we can reproduce the results, derived by Garibyan \cite{Garib} for the intensity, integrated over angles and frequencies $W^+ \sim \gamma$ and concentrated along the velocity of the particle with angular width $\theta \sim \gamma^{-1}$, and also, for the length of the radiation formation zone we get $\sim \gamma$, since the cutoff frequency $\omega_{\min} \sim \omega_p\gamma$, where $\omega_p$ is the plasma frequency, which characterizes the dispersion of the medium.

Then, we proceed in the following direction: suppose $N_s$ slabs of thickness $a_b$ and composed of a material with dielectric constant $b$ are immersed inside the medium with dielectric constant $\varepsilon$ and organized in a periodic structure with a fixed gap between two neighboring slabs $a_{\varepsilon}$. We are interested in the case $N_s \gg 1$, since in this case only those radiation fields are amplified, which correspond to the eigenmodes of the photonic crystal (PhC) with the same periodic structure. These modes are determined through the dispersion equation, see Supplemental Material.

\begin{gather}
    \cos(q_z l) = \cos(k_z^{\varepsilon} a_{\varepsilon})\cos(k_z^b a_b) - \nonumber\\ - \frac{1}{2} \left( \frac{\varepsilon^{-1}k_z^{\varepsilon}}{b^{-1}k_z^b} + \frac{b^{-1}k_z^b}{\varepsilon^{-1}k_z^{\varepsilon}} \right) \sin(k_z^{\varepsilon} a_{\varepsilon})\sin(k_z^b a_b),
    \label{dispeq}
\end{gather}

where $k_z^X$, $a_X$ determine the z-component of the wave-vector inside the medium $X = \varepsilon,b$ and the thickness of the slab, produced from the respective material and $l=a_{\varepsilon}+a_b$.

More precisely, the resonance condition here can be expressed as follows:

\begin{gather}
    \frac{\omega}{v} \mp q_z(\omega,q) \mod 2\pi/l = 0.
    \label{rescond}
\end{gather}

It means, that the particle has a wavevector respective to a given frequency and it excites only those eigenmodes of the photonic crystal, which match with its quasimomentum, up to an integer multiplier of the reciprocal lattice period.  
The larger $N_s$ we take, the narrower the peak becomes and the more precise match is required for amplification. But we are more interested in its neighborhood, rather than the peak itself, as explained below.

Note that at Brewster's angle $\varepsilon^{-1}k_z^{\varepsilon}=b^{-1}k_z^b$, therefore the rhs. of Eq.(\ref{dispeq}) becomes a  $\cos\left(k_z^{\varepsilon}a_{\varepsilon}+ k_z^b a_b\right)$, which means that there is always an eigenmode for the PhC composed of two homogeneous dielectrics, corresponding to the Brewster's transversal momentum $q_{Br} = \omega/c \times 1/\sqrt{b^{-1}+\varepsilon^{-1}}$. This natural opportunity suggests us to focus our attention on the resonance transition radiation, emitted along that angle.

For ultrarelativistic particle detectors the resonance condition is required to be fulfilled for a range of frequencies. It is shown in the Supplemental Material, that the condition above is satisfied once the PhC is designed according to 

\begin{gather}
    \frac{a_b}{a_{\varepsilon}}=\frac{\sqrt{\varepsilon+b}-\varepsilon}{b-\sqrt{\varepsilon+b}}.
    \label{idealprop}
\end{gather}

With increasing number of periods $N_s$ the sharper becomes the peak, which leads to the higher sensitivity of the observed intensity over even smaller deviations, induced by large Lorentz factor of the ultrarelativistic particle, from the resonance condition in Eq.(\ref{rescond}).

\textit{Results}- First of all, one of the ways to achieve powerful transition radiation is to choose two materials with strongly differing dielectric constants. However, in case when the PhC is composed of the materials, which differ significantly from the environment through their optical properties, this can prevent radiation propagating inside the PhC with Brewster's transversal momentum from escaping the system, once the following inequality $\sqrt{b^{-1}+\varepsilon^{-1}} \geq n^{-1}_{env}$ is violated ($n_{env}$ is the refractive index of the environment). If we choose vacuum as an environment and one of the slabs with the highest dielectric constant to be made of Si $b \approx 3.4^2$, then the second material should satisfy $\varepsilon < 1.05$, which is too close to the properties of vacuum. Thus, it becomes feasible to consider the system "identical, optically dense slabs - environment" from the beginning.

\begin{figure}
    \begin{tabular}{c}
        \begin{overpic}[width=0.9\columnwidth,keepaspectratio]{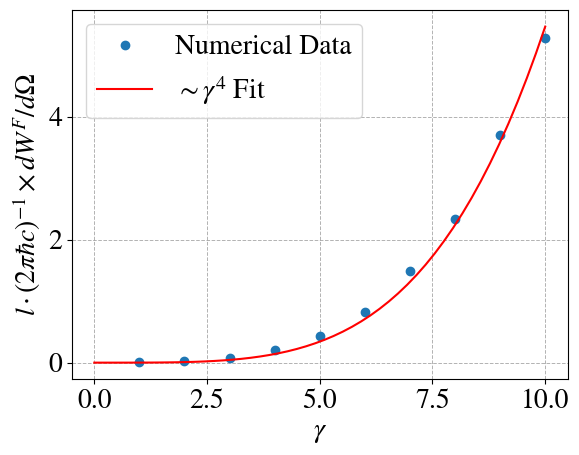}
        \put(-4,0.56\columnwidth){(a)} \end{overpic} \\
        \begin{overpic}[width=0.9\columnwidth,keepaspectratio]{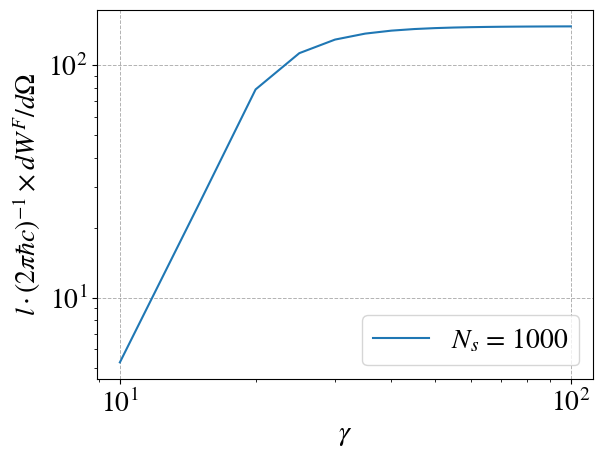}
        \put(-4,0.56\columnwidth){(b)} \end{overpic}
    \end{tabular}
    \caption{(a) Illustration of the $\sim \gamma^4$ law, (b) demonstration of the saturation law for the forward-propagating radiation. The spectral-angular intensity was integrated over the frequencies, corresponding to $\lambda > 2l$. Here $N_s = 10^3$.}
    \label{gamma4sat}
\end{figure}

Our proposal is to use the photonic crystal, composed of two non-dispersive materials with a particular structure inside a single period, given through the fraction of slab thickness and the gap between them according to Eq.(\ref{idealprop}). With this, the system now has a potential to expose a strong $\gamma$-dependence. The Spectral-angular intensity, observed at Brewster's angle along the forward direction:

\begin{gather}
    I^{+}(\omega,\theta_{Br}) \sim \Big| \frac{1-exp(iN_s\omega l\gamma^{-2}/2c)}{1-exp(i\omega l\gamma^{-2}/2c)} \Big|^2
\end{gather}

achieves its peak as $\gamma \rightarrow \infty$. In Fig.\ref{gamma4sat} it is demonstrated through the behavior of the angular intensity, obtained after the integration of the spectral-angular intensity over the spectrum of wavelengths $\lambda > 2(a_b+a_{\varepsilon}) = 2l$, denoted by ${\bf I}^+ = dW^F/d\Omega$. From $\gamma \sim \sqrt{N_s}$ it achieves the saturation peak $\sim N_s^2$ (just as in the frame of the Resonance TR effect \cite{TerMik}) and experiences no changes with the further increase of the particle energy, see Fig.\ref{gamma4sat}(b). But way below the saturation threshold, though where $\gamma \gg 1$ still holds, the integrated intensity exposes ${\bf I}^+(\theta = \theta_{Br}) \sim \gamma^4$ dependence, see Fig.\ref{gamma4sat}(a).

Note, that the strong energy dependence and the peak radiation intensity at Brewster's angle have been observed in early experiments \cite{gevexp}.

The dispersion equation Eq.(\ref{dispeq}) determines the range of frequencies, which are able to propagate along the PhC (allowed modes) and which are not (forbidden modes). Apparently, at Brewster's angle there are no forbidden modes (or no band gap) at any frequency. Once we determine the function $q_z = q_z(\omega,q) \in [0,\pi l^{-1}] $, we would like to express the frequency over $q_z$. And this, in general, results in band structure, since for a given $q_z$ there could be multiple corresponding frequencies $\omega$, which can be expressed as $\omega_n(q_z,\theta)$ with $n = 0, 1, 2,...$.

At Brewster's angle and taking into account, that $\varepsilon a_{\varepsilon}+b a_b = l\sqrt{\varepsilon+b}$, we get $\cos(q_zl) = \cos(\omega l/c)$.

Solving it with respect to $q_z$, we get

\begin{align}
    q_z l = \left\{ \begin{array}{cc}
        \omega l/c-\pi \Big\lfloor \frac{\omega}{\pi c l^{-1}} \Big\rfloor & ,\text{ if } \Big\lfloor \frac{\omega}{\pi c l^{-1}} \Big\rfloor = \text{even}, \\
        \pi \left( \Big\lfloor \frac{\omega}{\pi c l^{-1}} \Big\rfloor+1\right) - \omega l/c & ,\text{ if } \Big\lfloor \frac{\omega}{\pi c l^{-1}} \Big\rfloor = \text{odd},
    \end{array} \right.
\end{align}

which suggests, after denoting $n = \lfloor \omega/\pi c l^{-1} \rfloor$ and solving the equation with respect to $\omega$ for even and odd values of $n$ separately, the way to get the dispersion law: 

\begin{gather}
    \omega_n(q_z) = 2\pi c l^{-1} \Big\lfloor \frac{n+1}{2} \Big\rfloor + (-1)^n c \cdot sign(q_z) \times q_z,
\end{gather}

where $n = 0,1,2,...$ and we require the quasi-momentum $q_z \in (-\pi l^{-1},\pi l^{-1}]$, where the negative values emerge from the fact, that the dispersion equation is symmetric with respect to the transformation $q_z \rightarrow - q_z$. Even small deviations from the Brewster's angle cause an emergence of a gap in a band structure.

\begin{figure}[H]
    \begin{tabular}{c}
        \begin{overpic}[width=0.9\columnwidth,keepaspectratio]{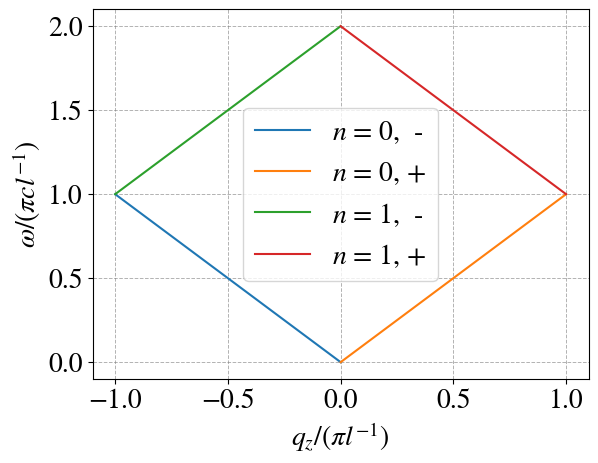}
        \put(0,0.56\columnwidth){(a)}
        \end{overpic} \\
        \begin{overpic}[width=0.9\columnwidth,keepaspectratio]{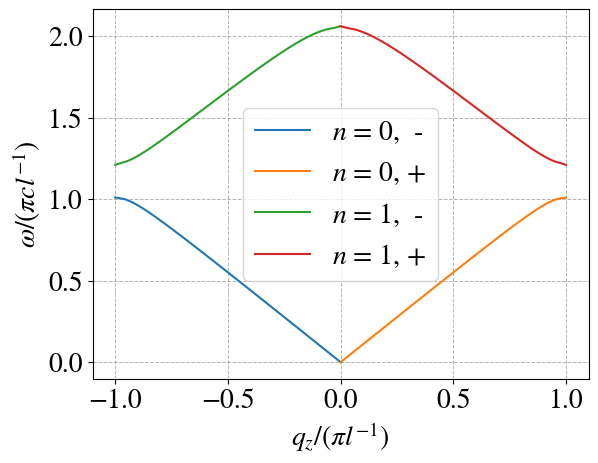}
        \put(0,0.56\columnwidth){(b)}
        \end{overpic}
    \end{tabular}
    \caption{Band structure of the described PhC for the modes propagating at (a) $\theta = \theta_{Br}$ and (b) $\theta = 1.013\cdot \theta_{Br}$. For labels, the number indicates $n$ for a respective branch, and the symbol following the number indicates the sign of $q_z$}
    \label{diraclike}
\end{figure}

With this, the band structure of the photonic crystal for the modes, propagating at Brewster's angle, resembles the Graphene Band structure near Dirac points; see Fig.\ref{diraclike}(a). Deviations from Brewster's angle play a role, analogous to the hopping parameter between the graphene layers \cite{Bern2013}; see Fig.\ref{diraclike}(b). The first number indicates $n$ for a respective branch, and the symbol following the number indicates the sign of $q_z$.

\textit{Discussion}- For ultrarelativistic particles, this radiation formation zone can increase significantly, leading to the distortions in observed intensity, since the mixed component of the Pointing vector starts to give a non-vanishing contribution. The structure we propose can eliminate this issue, present in X-ray transition radiation detectors (TRD).

The regular structures possess an objective advantage over the irregular ones in the X-ray region, as proved in \cite{Garib}. Besides, the formulas for the regular structure could be considerably simplified, while for the irregular one it can be done only in special cases. For example, \cite{pnas2025} considers a structure called "Brewster randomness", which allows the transition radiation from each interface to interfere constructively to give a sharp peak $\sim N_s^2$, with the angular width $\sim N_s^{-1}$, where $N_s$ is the number of slab pairs in that alternating structure. Second, the dependence of the radiation intensity was not considered, in contrary to the article \cite{Lin}, where photonic crystals (PhC) were considered and where sharp peaks of the spectral angular intensity were demonstrated along the isofrequency curves, which allowed to obtain a considerable $\gamma$ dependence of the intensity peak angle and, hence, the potential use in modern TRDs. For ultrarelativistic particles, the only problem was to design a photonic crystal in such a way, that in the given range of velocities of interest the respective eigenmodes were generated in the PhC and which showed a considerable angular correlation to $\gamma$.

It appears that the observation along the Brewster's angle has considerable advantages over those along $\theta = 0$. First, any photonic crystal has an eigenmode in that direction of propagation. Second, the quasi-momentum of that eigenmode is reduced to a linear combination of the crystal parameters. This allows to seek for an optimal design of the latter in order to achieve strong $\gamma$ dependence of the observables.

\begin{figure}[H]
    \begin{tabular}{c}
        \begin{overpic}[width=0.9\columnwidth,keepaspectratio]{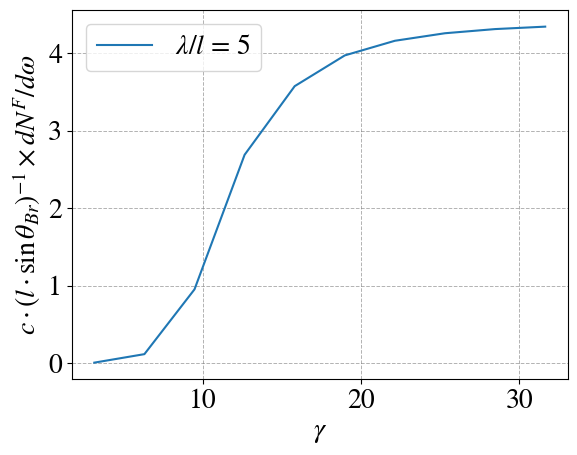}
        \put(-4,0.56\columnwidth){(a)} \end{overpic} \\
        \begin{overpic}[width=0.9\columnwidth,keepaspectratio]{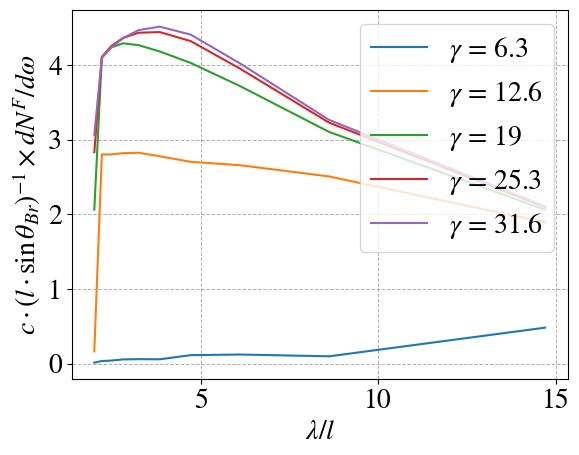}
        \put(-4,0.56\columnwidth){(b)} \end{overpic}
    \end{tabular}
    \caption{Differential yield for the photons, radiated along the Brewster's angle in forward direction, integrated over a window $|\delta\theta| < 4 \cdot 10^{-3}$, (a) illustration of $\gamma$-dependence for a fixed wavelength $\lambda = 5l$, (b) illustration of $\lambda$-dependence for various fixed values of $\gamma$. Here $N_s = 10^3$.}
    \label{IntYield}
\end{figure}

According to the numbers in Fig. \ref{IntYield}, the Yield of photons is typically $\sim 0.1 - 1 eV^{-1}$ for the PhC period $l \sim 1\mu m$, which is much higher compared to that of the X-ray TRD \cite{AndWes}. This is due to the much lower energies for the single photons in the vicinity of the visible region. Additionally, in contrast to TRD, the radiation formation zone remains limited to $\lambda/2|1-\varepsilon/\sqrt{\varepsilon+b}|$ in the ultrarelativistic limit. This makes another advantage over the conventional X-ray TRD.

The perspective of the detectors for registering particles with $\gamma \sim 10^3 - 10^4$ seems problematic to realize with the state-of-the-art technologies, since it would require the manufacture of periodic nanostructures with $N_s \sim 10^6-10^8$. However, some highly advancing methods like Block Copolymer Lithography (BCP) already allow achieving and even far exceeding the described period number \cite{Miri1997,Cheng2025}. Thus, the only problem that remains is the precise angular spectroscopy, which will allow restricting the radiation to pass through an enough narrow range of angles $\sim 0.1$ arcseconds. Some articles suggest the possibility of angular measurements with an error $\sim 1$ nanoradians, i. e. with even more accuracy \cite{pisani2006}.

\textit{Summary}- We considered the transition radiation, induced by a particle traversing perpendicular to the layers of the stratified medium. In particular case of the bimaterial photonic crystal with a specific proportion held for thicknesses of its components, the emitted radiation becomes highly dependent on the particle energy, which is absent in case of the other mechanisms, like Cherenkov radiation, bremsstrahlung, luminescence and scintillation. Along with the state of the art nanotechnologies and precision angular spectroscopy, we have the potential to manufacture compact transition radiation detectors, operating near the optical region for the detection of the ultrarelativistic energies. 


\textit{Acknowledgement}- This work was supported by the Higher Education and Science Committee of MESCS RA, in the frame of the research projects No. 21AG-1C062 and No. 23-2DP-1C010.


\end{document}


\title{- Supplemental Material -\\ Strong Energy Dependent Transition Radiation in a Photonic Crystal}
\author{V.Gareyan$^{1}$ and Zh.Gevorkian$^{1,2}$}
\affiliation{$^{1}$ Alikhanyan National Laboratory, Alikhanian Brothers St. 2,  Yerevan 0036 Armenia\\
$^{2}$ Institute of Radiophysics and Electronics, Ashtarak-2 0203 Armenia}

\maketitle

\tableofcontents

\renewcommand{\theequation}{S.\arabic{equation}}
\setcounter{equation}{0}
\renewcommand{\thefigure}{S.\arabic{figure}}
\setcounter{figure}{0}

\section{Note 1: Theory}

\subsection{1.1 Initial considerations}
\subsubsection{1.1.1 Wave equation with a source}



We start from the Maxwell's equations ($\mu = 1$):

\begin{align}
    \begin{array}{c}
        \nabla \times {\bf E} = -\frac{1}{c} \frac{\partial {\bf H}}{\partial t} \\
        \nabla \times {\bf H} = \frac{4\pi}{c} ({\bf j}_{c}+{\bf j}_{\sigma}) + \frac{1}{c} \frac{\partial {\bf D}}{\partial t} \\
        \nabla \cdot {\bf D} = 4\pi (\rho_{c} + \rho_{\sigma}) \\
        \nabla \cdot {\bf H} = 0
    \end{array}
\end{align}

where all fields are

\begin{eqnarray}
    {\bf F} = {\bf F} ({\bf r}, t) \nonumber
\end{eqnarray}

and ${\bf j}_c$ and $\rho_c$ denote the uncompensated electric current and charge densities outside the material, respectively, and ${\bf j}_{\sigma}$ and $\rho_{\sigma}$ are those for the uncompensated charges inside the material. In our case of a single electron, moving uniformly with velocity ${\bf v} = {\bf e}_z*v$:

\begin{align}
    \begin{array}{c}
        {\bf j}_c({\bf r}, t) = e{\bf v} \delta^{(3)}({\bf r} - {\bf v} t) \\
        \rho_c({\bf r}, t) = e \delta^{(3)}({\bf r} - {\bf v} t)
    \end{array}
\end{align}

By performing a Fourier transform on t according to:

\begin{eqnarray}
    {\bf F}({\bf r}, t) = \int_{-\infty}^{\infty} \frac{d\omega}{2\pi} e^{-i\omega t} {\bf F}({\bf r}, \omega)
\end{eqnarray}

we arrive at 

\begin{align}
    \begin{array}{c}
        \nabla \times {\bf E} = i \frac{\omega}{c} {\bf H} \\
        \nabla \times {\bf H} = \frac{4\pi}{c} ({\bf j}_{c}+\sigma {\bf E}) -i \frac{\omega}{c} {\bf D} \\
        \nabla \cdot {\bf D} = 4\pi (\rho_{c}+\rho_{\sigma}) \\
        \nabla \cdot {\bf H} = 0
    \end{array}
\end{align}

where now 

\begin{align}
    \begin{array}{c}
        {\bf j}_c({\bf r}, \omega) = e{\bf v} \int_{-\infty}^{\infty} dt e^{+i\omega t} \delta^{(3)}({\bf r} - {\bf v} t) = {\bf e}_z*e\delta^{(2)}(\boldsymbol{\rho})e^{i\frac{\omega}{v}z} \\
        \rho_c({\bf r}, \omega) = \frac{e}{v} \delta^{(2)}(\boldsymbol{\rho})e^{i\frac{\omega}{v}z}
    \end{array}
\end{align}

Here we used ${\bf j}_{\sigma} = \sigma {\bf E}$ and the following properties of the $\delta$-functions:

\begin{eqnarray}
    \delta^{(3)}({\bf r} - {\bf v}t)=\delta^{(2)}(\boldsymbol{\rho})*\delta(z - vt) = \nonumber\\  =\delta^{(2)}(\boldsymbol{\rho})*\frac{1}{v}\delta(t-\frac{z}{v})
\end{eqnarray}

In the frequency domain, we can use $ {\bf D}({\bf r}, \omega) = \varepsilon_d (z, \omega){\bf E}({\bf r}, \omega)$, where the dielectric constant of our slab system:

\begin{eqnarray}
    \varepsilon_d (z, \omega) = \sum_{i=0}^{N+1} \varepsilon_{d,i} (\omega)*[\Theta(z-z_{i-1})-\Theta(z-z_i)]
\end{eqnarray}

where $z_{-1} = -L_z/2$, $z_i, i=\overline{0,N}$ are the points of contact of the media with dielectric coefficients $\varepsilon_{d,i}$ and $\varepsilon_{d,i+1}$, $z_{N+1} = +L_z/2$ and $\Theta(z)$ is the Heaviside step function.

Using the continuity equation for the total uncompensated charges, and taking into account, that that equation is satisfied separately for the uniformly moving charge, we get $\rho_c=0$. With this we can write

\begin{align}
    \begin{array}{c}
        \nabla \times {\bf E} = i \frac{\omega}{c} {\bf H} \\
        \nabla \times {\bf H} = \frac{4\pi}{c} {\bf j}_{c} -i \frac{\omega}{c} \varepsilon {\bf E} \\
        \nabla \cdot {\bf D} = 4\pi \rho_{c} \\
        \nabla \cdot {\bf H} = 0
    \end{array}
\end{align}

where $\varepsilon(\omega) = \varepsilon_d(\omega) + i4\pi\sigma(\omega)/\omega$ is the effective dielectric constant. Thus, the system behaves as a non-conductor with the effective dielectric constant

\begin{eqnarray}
    \varepsilon (z, \omega) = \sum_{i=0}^{N+1} \varepsilon_{i} (\omega)*[\Theta(z-z_{i-1})-\Theta(z-z_i)]
\end{eqnarray}

Note that because for all fields ${\bf F}({\bf r}, -\omega) = {\bf F}^*({\bf r}, \omega)$, we have $\varepsilon(z, -\omega) = \varepsilon^*(z, \omega)$. This fact will be used later.

Because of axial symmetry of our system with respect to the z-axis, we can go further at Fourier decomposition in tangential space:

\begin{eqnarray}
    {\bf F}({\bf r}, \omega) = \int \int \frac{d^2{\bf q}}{(2\pi)^2} e^{i{\bf q} \cdot \boldsymbol{\rho}} {\bf F}(z, {\bf q}, \omega)
\end{eqnarray}

where we finally get the following for the Maxwell equations:

\begin{align}
    \begin{array}{c}
        (i{\bf q} + {\bf e}_z*\partial_z) \times {\bf E} = i \frac{\omega}{c} {\bf H} \\
        (i{\bf q} + {\bf e}_z*\partial_z) \times {\bf H} = \frac{4\pi}{c} {\bf j}_{c} -i \frac{\omega}{c} \varepsilon {\bf E} \\
        (i{\bf q} + {\bf e}_z*\partial_z) \cdot {\bf D} = 4\pi \rho_{c} \\
        (i{\bf q} + {\bf e}_z*\partial_z) \cdot {\bf H} = 0
    \end{array}
    \label{maxwellfinal}
\end{align}

with 

\begin{align}
    \begin{array}{c}
        {\bf j}_c(z, {\bf q}, \omega) = {\bf e}_z*e\int \int d^2\boldsymbol{\rho} e^{-i{\bf q}\cdot \boldsymbol{\rho}} \delta^{(2)}(\boldsymbol{\rho})e^{i\frac{\omega}{v}z} = {\bf e}_z*e* e^{i\frac{\omega}{v}z}\\
        \rho_c(z, {\bf q}, \omega) = \frac{e}{v} e^{i\frac{\omega}{v}z}
    \end{array}
\end{align}

Performing by $(i{\bf q} + {\bf e}_z \partial_z) \times$ on the second equation in system (\ref{maxwellfinal}) and making use of all the remaining formulas of the system, we get:

\begin{eqnarray}
    \left[ \frac{\omega^2}{c^2}\varepsilon(z, \omega) - q^2 +\partial_z^2 \right] {\bf H} = -\frac{4\pi i e}{c} ({\bf q} \times {\bf e}_z) e^{i\frac{\omega}{v}z} + \nonumber\\ + i\frac{\omega}{c} (\partial_z \varepsilon) ({\bf e}_z \times {\bf E})
\end{eqnarray}

But since it is enough for us to consider the entire domain $z \in (-L_z/2, L_z/2)$ excluding contact points $z_i$, which are the only places where $\partial_z\varepsilon \ne 0$, we finally get the wave equation: 

\begin{eqnarray}
    \left[ \frac{\omega^2}{c^2}\varepsilon(z, \omega) - q^2 +\partial_z^2 \right] {\bf H} = -\frac{4\pi i e}{c} ({\bf q} \times {\bf e}_z) e^{i\frac{\omega}{v}z} \equiv \nonumber\\ \equiv {\bf J}(z, {\bf q}, \omega)
\end{eqnarray}

We can convert this equation to a 1D form because the source term here is collinear to the unit vector ${\bf e}_{\Phi} = ({\bf e}_z \times \hat{q})$. This implies that the only projection of the magnetic field vector that survives is its projection along exactly that vector:

\begin{eqnarray}
    \left[ \frac{\omega^2}{c^2}\varepsilon(z, \omega) - q^2 +\partial_z^2 \right] H_{\Phi} = \frac{4\pi i e q}{c} e^{i\frac{\omega}{v}z}
    \label{waveeqH}
\end{eqnarray}

\subsubsection{1.1.2 Boundary Conditions and the Solution}

We can find a general solution as a sum of a partial solution of the inhomogeneous equation and the general solution of the corresponding homogeneous one. Starting with the partial solution, we note that the source in the RHS. of (\ref{waveeqH}) is $\sim exp(iz\omega/v)$, suggesting that one searches it in the respective auxiliary form:

\begin{eqnarray}
    H^e_{\Phi} = \frac{4\pi i e q}{c} \frac{e^{i\frac{\omega}{v}z}}{\omega^2*(\varepsilon(z, \omega)/c^2 - 1/v^2) - q^2}
\end{eqnarray}

Note that this partial solution is correct only if we exclude contact points $z_i$ from the consideration domain. We need to add the general form of the solution of (\ref{waveeqH}) without any sources:

\begin{eqnarray}
    H^r_{\Phi} = \sum_{i=0}^{N+1}[\Theta(z-z_{i-1})-\Theta(z-z_i)] * \nonumber\\ * [A_i^+ e^{i k_{i,z}(z-z_i)} + A_i^- e^{-i k_{i,z}(z-z_i)}],
    \label{radH}
\end{eqnarray}

where $k_{i,z} = sgn(\omega)\cdot \sqrt{(\omega/c)^2\varepsilon_i(\omega)-q^2}$. Such a definition is required to ensure the reality of the fields ${\bf F}({\bf r}, t)$.

The coefficients $A_i^{\pm}$ are determined from the need for electric and magnetic fields to satisfy the boundary conditions at the contact points $z_i, i = \overline{0,N}$.

Before proceeding, we need to make an important note: Since ${\bf H}(z, -{\bf q}, -\omega) = {\bf H}^*(z, {\bf q}, \omega)$ and ${\bf e}_{\Phi}(-\hat{q}) = - {\bf e}_{\Phi}(\hat{q})$, we have $H_{\Phi}(z, -{\bf q}, -\omega) = - H_{\Phi}^*(z, {\bf q}, \omega)$. In addition to the properties of the dielectric constant $\varepsilon=\varepsilon^{\prime}+ i\varepsilon^{\prime \prime}$, in the case of small absorptions

\begin{gather}
    k_{z}(\omega, q) \approx \frac{\omega}{c} \sqrt{\varepsilon^{\prime}(\omega) - \frac{c^2q^2}{\omega^2}}\cdot\left( 1 + i \frac{\varepsilon^{\prime \prime}(\omega)}{\varepsilon^{\prime}(\omega) - c^2q^2/\omega^2} \right) = \nonumber\\ = \frac{\omega}{c} \left( \sqrt{\varepsilon^{\prime}(\omega) - \frac{c^2q^2}{\omega^2}} + i \frac{\varepsilon^{\prime \prime}(\omega)}{2\sqrt{\varepsilon^{\prime}(\omega) - c^2q^2/\omega^2}}\right) \nonumber\\ \Theta \left(\sqrt{\varepsilon^{\prime}}\frac{|\omega|}{c} - q \right) + \nonumber\\ + \frac{\omega}{c} \left( i\sqrt{\frac{c^2q^2}{\omega^2} - \varepsilon^{\prime}(\omega)} +  \frac{\varepsilon^{\prime \prime}(\omega)}{2\sqrt{c^2q^2/\omega^2 - \varepsilon^{\prime}(\omega)}}\right) \nonumber\\ \Theta \left(q - \sqrt{\varepsilon^{\prime}}\frac{|\omega|}{c}\right)
\end{gather}

(we omitted the subindex of the media number) we have:

\begin{gather}
    k_{i,z}(-\omega, q) = - k_{i,z}^*(\omega, q) \Theta\left(\sqrt{\varepsilon_i^{\prime}}|\omega|/c - q\right) + \nonumber\\ k_{i,z}^*(\omega, q) \Theta\left(q - \sqrt{\varepsilon_i^{\prime}}|\omega|/c \right)
\end{gather}

Using this, after separating the linearly independent exponents in \ref{radH}, we obtain the following for the amplitudes of the magnetic field:

\begin{gather}
    A_i^{+*} (\omega, {\bf q}) = - A_i^+(-\omega,-{\bf q})\Theta\left(\sqrt{\varepsilon_i^{\prime}}|\omega|/c - q \right) -\nonumber\\ - A_i^-(-\omega,-{\bf q})\Theta\left(q - \sqrt{\varepsilon_i^{\prime}}|\omega|/c \right), \nonumber\\
    A_i^{-*} (\omega, {\bf q}) = - A_i^-(-\omega,-{\bf q})\Theta\left(\sqrt{\varepsilon_i^{\prime}}|\omega|/c - q\right) -\nonumber\\ - A_i^+(-\omega,-{\bf q})\Theta \left(q - \sqrt{\varepsilon_i^{\prime}}|\omega|/c\right).
\end{gather}

This implies, in particular, that on the right and leftmost media only the components with $q \leq \sqrt{\varepsilon^{\prime}}|\omega|/c$ survive.

Because 

\begin{eqnarray}
    {\bf E} = \frac{c}{i\omega \varepsilon(z,\omega)} \left[ {\bf e}_z \left( \frac{4\pi}{c} j_z -iqH_{\Phi} \right)  + \hat{q} \partial_zH_{\Phi}\right]
    \label{ElectricField}
\end{eqnarray}

we find that the quantities $H_{\Phi} \text{ and } \varepsilon^{-1}\partial_z H_{\Phi}$ need to be continuous. We can write this mathematically as a system of equations:

\begin{gather}
    H_{\Phi}(z_j^-) = H_{\Phi}(z_j^+) \nonumber\\
    \varepsilon^{-1}\partial_z H_{\Phi}(z_j^-) = \varepsilon^{-1}\partial_z H_{\Phi} (z_j^+)
\end{gather}

Since 

\begin{eqnarray}
    \Theta(z_j^- -z_{i-1})-\Theta(z_j^- -z_i) = \delta_{ij} \nonumber\\
    \Theta(z_j^+ -z_{i-1})-\Theta(z_j^+ -z_i) = \delta_{i,j+1}
\end{eqnarray}

we can write this in matrix form:

\begin{gather}
    \left(\begin{array}{c}
        A_j^+ \\
        A_j^-
    \end{array}\right) = \hat{B}_j + \hat{M}_j \left(\begin{array}{c}
        A_{j+1}^+ \\
        A_{j+1}^-
    \end{array}\right), \text{ } \forall j = \overline{0,N}
    \label{recurrent}
\end{gather}

with

\begin{widetext}
\begin{gather}
    \hat{B}_j = \frac{2\pi i e q}{c k_{j,z} \varepsilon^{-1}_j} e^{i z_j \omega/v} \left( \begin{array}{c}
        \left( \frac{\varepsilon^{-1}_j k_{j,z} + \varepsilon^{-1}_{j+1} \omega/v}{k_{j+1,z}^2 - (\omega/v)^2} - \frac{\varepsilon^{-1}_{j}}{k_{j,z} - \omega/v} \right) \\
        \left( \frac{\varepsilon^{-1}_j k_{j,z} - \varepsilon^{-1}_{j+1} \omega/v}{k_{j+1,z}^2 - (\omega/v)^2} - \frac{\varepsilon^{-1}_{j}}{k_{j,z} + \omega/v} \right)
    \end{array}\right) \nonumber\\
    \hat{M}_{j} = \frac{1}{2k_{j,z}\varepsilon_j^{-1}} \left( \begin{array}{cc}
        e^{-ik_{j+1,z}(z_{j+1}-z_j)}(k_{j,z}\varepsilon_j^{-1}+k_{j+1,z}\varepsilon_{j+1}^{-1}) & e^{ik_{j+1,z}(z_{j+1}-z_j)}(k_{j,z}\varepsilon_j^{-1}-k_{j+1,z}\varepsilon_{j+1}^{-1}) \\
        e^{-ik_{j+1,z}(z_{j+1}-z_j)}(k_{j,z}\varepsilon_j^{-1}-k_{j+1,z}\varepsilon_{j+1}^{-1}) & e^{ik_{j+1,z}(z_{j+1}-z_j)}(k_{j,z}\varepsilon_j^{-1}+k_{j+1,z}\varepsilon_{j+1}^{-1})
    \end{array} \right)
\end{gather}
\end{widetext}

Applying the recurrent relation (\ref{recurrent}) for the column in the RHS over and over, we finally get:

\begin{gather}
    \left(\begin{array}{c}
        A_j^+ \\
        A_j^-
    \end{array}\right) = \sum_{i=j}^{N} \left[ \prod_{k=j}^{i-1} \hat{M}_k \right] \hat{B}_i + \prod_{k=j}^{N} \hat{M}_k \left(\begin{array}{c}
        A_{N+1}^+ \\
        0
    \end{array}\right)
\end{gather}

with the use of the fact of the absence of backpropagation in the right-most medium. For $j=0$, we use the absence of forward propagating modes in the left-most medium to obtain the following.

\begin{eqnarray}
    A_{N+1}^+ = - B_1/M_{11} \text{ and } A_0^- = B_2 - \frac{M_{21}}{M_{11}} B_1 \nonumber\\
    \text{with } \hat{B} = \sum_{i=0}^{N} \left[ \prod_{k=0}^{i-1} \hat{M}_k \right] \hat{B}_i \text{ and } \hat{M} = \prod_{k=0}^{N} \hat{M}_k
    \label{rightleftmost}
\end{eqnarray}

Consider a particular case of a stratified system. For a system of $N_s$ stacks with the dielectric constant $b$ (we take the imaginary part to be zero for any dielectric constant) and arbitrary thicknesses and arbitrary distances between each other, inside the medium with the dielectric constant $\varepsilon$ we have the following. 


\begin{gather}
    \varepsilon_k = \varepsilon \cdot \delta_{0, k \mod 2} + b \cdot \delta_{1, k \mod 2}, \forall k = \overline{0, 2N_s} 
\end{gather}

With this

\begin{gather}
    \hat{B} = \sum_{j=0}^{N_s-1} \left[ \prod_{k=0}^{j-1} \hat{m}_k \right] \hat{C}_{j} , 
    \nonumber\\ \hat{M} = \prod_{k=0}^{N_s-1} \hat{m}_k, \nonumber\\ \text{ where } \hat{m}_k = \hat{M}_{2k}\hat{M}_{2k+1} = \left( \begin{array}{cc}
        (m_k)_{11} & (m_k)_{12} \\
        (m_k)_{12}^* & (m_k)_{11}^*
    \end{array} \right), \nonumber\\
    \hat{C}_{j} = \hat{B}_{2j} + \hat{M}_{2j} \hat{B}_{2j+1} = \left( \begin{array}{c}
        (C_j)_1 \\
        (C_j)_2
    \end{array} \right)
\end{gather}

with

\begin{widetext}
\begin{gather}
    (m_k)_{11} = e^{-ik_z^{\varepsilon}d_{2k+2}}\left[\cos{(k_z^b d_{2k+1})}-\frac{i}{2} \left( \frac{{\varepsilon}^{-1}k_z^{\varepsilon}}{b^{-1}k_z^b} + \frac{b^{-1}k_z^b}{{\varepsilon}^{-1}k_z^{\varepsilon}} \right) \sin{(k_z^bd_{2k+1})}\right], \nonumber\\
    (m_k)_{12} = e^{ik_z^{\varepsilon}d_{2k+2}} \frac{i}{2} \left( \frac{{\varepsilon}^{-1}k_z^{\varepsilon}}{b^{-1}k_z^b} - \frac{b^{-1}k_z^b}{{\varepsilon}^{-1}k_z^{\varepsilon}} \right) \sin{(k_z^bd_{2k+1})}, \nonumber\\
    (C_j)_1 = \frac{2\pi i eq}{c} e^{iz_{2j} \omega/v} \left[ \frac{1}{(k_z^b)^2-\omega^2/v^2} \left( \left[ 1+\frac{b^{-1}\omega/v}{\varepsilon^{-1}k_z^{\varepsilon}} \right] \left[ 1-e^{id_{2j+1}\omega/v}\cos{(k_z^bd_{2j+1})} \right] + i e^{id_{2j+1}\omega/v} \left[ \frac{b^{-1}k_z^b}{{\varepsilon}^{-1}k_z^{\varepsilon}}+ \frac{\omega/v}{k_z^b}  \right] \sin{(k_z^bd_{2j+1})} \right) \right. \nonumber\\ \left. - \frac{1}{(k_z^{\varepsilon})^2-\omega^2/v^2} \left( \left[ 1+\frac{\omega/v}{k_z^{\varepsilon}} \right] \left[ 1-e^{id_{2j+1}\omega/v}\cos{(k_z^bd_{2j+1})} \right] + i e^{id_{2j+1}\omega/v} \left[ \frac{b^{-1}k_z^b}{{\varepsilon}^{-1}k_z^{\varepsilon}} + \frac{\varepsilon^{-1}\omega/v}{b^{-1}k_z^b}  \right] \sin{(k_z^bd_{2j+1})} \right) \right] ,\nonumber\\
    (C_j)_2 = \frac{2\pi i eq}{c} e^{iz_{2j} \omega/v} \left[ \frac{1}{(k_z^b)^2-\omega^2/v^2} \left( \left[ 1-\frac{b^{-1}\omega/v}{\varepsilon^{-1}k_z^{\varepsilon}} \right] \left[ 1-e^{id_{2j+1}\omega/v}\cos{(k_z^bd_{2j+1})} \right] - i e^{id_{2j+1}\omega/v} \left[ \frac{b^{-1}k_z^b}{{\varepsilon}^{-1}k_z^{\varepsilon}} - \frac{\omega/v}{k_z^b}  \right] \sin{(k_z^bd_{2j+1})} \right) \right. \nonumber\\ \left. - \frac{1}{(k_z^{\varepsilon})^2-\omega^2/v^2} \left( \left[ 1-\frac{\omega/v}{k_z^{\varepsilon}} \right] \left[ 1-e^{id_{2j+1}\omega/v}\cos{(k_z^bd_{2j+1})} \right] - i e^{id_{2j+1}\omega/v} \left[ \frac{b^{-1}k_z^b}{{\varepsilon}^{-1}k_z^{\varepsilon}} - \frac{\varepsilon^{-1}\omega/v}{b^{-1}k_z^b}  \right] \sin{(k_z^bd_{2j+1})} \right) \right],
    \label{matrixEq}
\end{gather}
\end{widetext}

where $d_i = z_i - z_{i-1}$ is the thickness of the (i-1)-th slab.

\subsubsection{1.1.3 Derivation of the formula for the Spectral Angular Intensity}

Before starting the section, some preliminary judgments are required. First, in order to observe any radiation far enough from the given radiating system, the medium inside which the electromagnetic waves traverse should have a little enough absorption. We will start with the assumption that for any material the absorption, described by $\varepsilon_i^{\prime \prime} = \Im m(\varepsilon_i)$ is small enough to perform a respective Taylor decomposition, and at the end we will get a more specific mathematical form of that expression.

The Poynting vector 

\begin{eqnarray}
    {\bf S} ({\bf r}, t) = \frac{c}{4\pi} {\bf E} ({\bf r}, t) \times {\bf H} ({\bf r}, t)
    \label{poynting}
\end{eqnarray}

assists us in defining the energy transferred through a given surface $\Sigma$ with a definite normal ${\bf n} = d\boldsymbol{\Sigma}/d\Sigma$ everywhere on it. \color{black} Since we assume a negligible absorption of radiation, thus it just traverses through the matter without causing any changes inside it, the transferred energy should not depend on the surface that includes a given radiation source. \color{black}


\begin{gather}
     W = \int_{-\infty}^{+\infty} dt \int_{\Sigma} d\boldsymbol{\Sigma} \boldsymbol \cdot {\bf S} = \nonumber\\ = \frac{c}{4\pi} \int_{-\infty}^{+\infty} \frac{d\omega}{2\pi} \frac{d\omega'}{2\pi} \int_{-\infty}^{+\infty} dt e^{-i(\omega + \omega')t} \nonumber\\ \int_{\Sigma} d\boldsymbol{\Sigma} \cdot [{\bf E} ({\bf r}, \omega) \times {\bf H} ({\bf r}, \omega')]
\end{gather}

Since 

\begin{gather}
    \int_{-\infty}^{+\infty} dt e^{-i(\omega + \omega')t} = 2\pi \delta (\omega + \omega')
\end{gather}

we get

\begin{gather}
    W = \frac{c}{4\pi}   \int_{-\infty}^{+\infty} \frac{d\omega}{2\pi} \int_{\Sigma} d\boldsymbol{\Sigma} \cdot [{\bf E} ({\bf r}, \omega) \times {\bf H} ({\bf r}, -\omega)] = \nonumber\\ = \frac{c}{4\pi^2}   \int_{0}^{+\infty} d\omega \int_{\Sigma} d\boldsymbol{\Sigma} \cdot \Re[{\bf E} ({\bf r}, \omega) \times {\bf H} ({\bf r}, -\omega)]
\end{gather}

where we used the fact that the fields were real quantities: ${\bf F} ({\bf r}, t) = {\bf F}^* ({\bf r}, t)$. 

Because ${\bf H} = {\bf H}^e + {\bf H}^r$ and using (\ref{ElectricField}) we get the same decomposition for the electric field ${\bf E} = {\bf E}^e + {\bf E}^r$, we can write the total transmitted energy as the sum of the following 3 components \cite{Garib}: $W = W^{e} + W^{r} + W^{int}$, where

\begin{gather}
    W^{e} = \frac{c}{4\pi^2} \int_{0}^{+\infty} d\omega \int_{\Sigma} d\boldsymbol{\Sigma} \cdot \Re[{\bf E}^e ({\bf r}, \omega) \times {\bf H}^e ({\bf r}, -\omega)], \nonumber\\
    W^{r} = \frac{c}{4\pi^2} \int_{0}^{+\infty} d\omega \int_{\Sigma} d\boldsymbol{\Sigma} \cdot \Re[{\bf E}^r ({\bf r}, \omega) \times {\bf H}^r ({\bf r}, -\omega)], \nonumber\\
    W^{int} = \frac{c}{4\pi^2} \int_{0}^{+\infty} d\omega \int_{\Sigma} d\boldsymbol{\Sigma} \cdot \Re[{\bf E}^e ({\bf r}, \omega) \times {\bf H}^r ({\bf r}, -\omega) + \nonumber\\ + {\bf E}^r ({\bf r}, \omega) \times {\bf H}^e ({\bf r}, -\omega)].
    \label{poynComponents}
\end{gather}


The first component represents the Cherenkov radiation, once the particle velocity exceeds the phase velocity of light propagating in the medium. In the classical literature \cite{Jelley}, it can be calculated as an energy flow through the infinite cylinder, with its central axis directed along the trajectory of the particle. It can also be proved that there is no energy transfer from this component if the particle velocity is below the threshold. The formula is as follows.

\begin{gather}
    W^e = \frac{e^2}{c^2} \int_{-\infty}^{+\infty} dz \int_{0}^{+\infty}d\omega \omega \left( 1-\frac{1}{\beta^2\varepsilon^{\prime}(z, \omega)} \right) \nonumber\\ \cdot \Theta\left( \beta^2\varepsilon^{\prime}(z, \omega)-1 \right)
\end{gather}

where $\beta = v/c$.

Consider now the contribution of the transition radiation, i. e. the second term; in the rightmost and leftmost media choose $\Sigma$ as a plane, perpendicular to the z-axis, is inside the medium and with the normal ${\bf n} = {\bf e}_z$ and ${\bf n} = -{\bf e}_z$ respectively. Along with the Fourier decomposition with respect to the transverse momentum, using the identity

\begin{gather}
    \int d^2\boldsymbol{\rho} e^{i({\bf q} + {\bf q}')\boldsymbol{\rho}} = (2\pi)^2 \delta({\bf q} + {\bf q}').
\end{gather}

leads to

\begin{gather}
    W^{r,\pm} = \pm \frac{c}{4\pi^2} \int \frac{d^2 {\bf q}}{(2\pi)^2} \int_{0}^{ +\infty} d\omega \nonumber\\ \Re \left[ \frac{c}{i\omega \varepsilon(z, \omega)} H_{\Phi}^{r,*} \partial_z H_{\Phi}^r \right]
\end{gather}


Denote $\varepsilon_0 = \varepsilon_{-}$ and $\varepsilon_{N+1} = \varepsilon_{+}$ and the z-coordinates of the fixed observation point through $z^{obs}_{\pm}$ respectively to arrive to 

\begin{gather}
    W^{r,\pm} = \frac{c}{4\pi} \int \frac{d^2 {\bf q}}{(2\pi)^2} \int_{-\infty}^{\infty} \frac{d\omega}{2\pi} \frac{c}{\omega}  |H_{\Phi}^{r}(z^{obs}_{\pm}, {\bf q}, \omega)|^2 \nonumber\\ \cdot \Re\left[\frac{1}{\varepsilon_{\pm}}\sqrt{\frac{\omega^2}{c^2}\varepsilon_{\pm}-q^2}\right]
\end{gather}

As proved previously, the Fourier components of the magnetic field are non-zero only for $q < \omega \sqrt{\varepsilon^{\prime}_{\pm}}/c$, which allows us to convert the integration over momentum to the integration over the solid angle in the following way: Since the momentum inside each media can be represented as ${\bf q} = \sqrt{\varepsilon^{\prime}_{\pm}} \omega/c \cdot \sin{\theta_{\pm}} ({\bf e_x}\cos{\phi_{\pm}} + {\bf e_y}\sin{\phi_{\pm}})$ and since the solid angle is $d\Omega = \sin{\theta} d\theta d\phi$, we can write

\begin{eqnarray}
    \int_{q < \omega \sqrt{\varepsilon^{\prime}_{\pm}}/c} \frac{d^2 {\bf q}}{(2\pi)^2} = \left( \frac{\omega \sqrt{\varepsilon^{\prime}_{\pm}}}{2\pi c} \right)^2 \int_{(2\pi)^+} d\Omega_{\pm} \cos{\theta_{\pm}}
\end{eqnarray}

Note that for the reflected waves the count of the polar $\theta_-$ angle starts from the negative direction along the z-axis. This allows us to write the energy flux as an integral over the solid angle and all the possible frequencies:

\begin{gather}
    W^{r,\pm} = \int_{0}^{\infty} d\omega \int_{(2\pi)^+} d\Omega_{\pm} \nonumber\\ c \left( \frac{\omega}{4\pi^2 c} \right)^2 \sqrt{\varepsilon^{\prime}_{\pm}} \cos^2 \theta_{\pm} |A^{\pm}(\omega, \theta_{\pm})|^2 e^{\mp 2\Im[k_z^{\pm}](z^{obs}_{\pm}-z_{\pm})}.
\end{gather}

Note the exponential decay term. It is tailored to the absorption properties of the medium, while can be neglected if $|z^{obs}_{\pm}-z_{\pm}|\ll1/2\Im[k_z^{\pm}]$.

We omitted the subindex for each of the coefficients $A^{\pm}$, since it is apparent that the "+" sign refers to $A_{N+1}^+$ and the "-" sign refers to $A_0^-$. Thus, the spectral angular intensity:

\begin{eqnarray}
    I^{r,\pm}(\omega,\theta_{\pm}) = c \left( \frac{\omega}{4\pi^2 c} \right)^2 \sqrt{\varepsilon^{\prime}_{\pm}} \cos^2 \theta_{\pm} |A^{\pm}(\omega, \theta_{\pm})|^2 \nonumber\\
    \label{SpecAngI}
\end{eqnarray}

The third component represents the interference between the solutions of homogeneous and inhomogeneous solutions of the wave equation. It can be proved \cite{Garib}, that this component contains the exponential term $exp(-iz/z^{\pm}(\theta))$, where $z^{\pm}(\theta) = \pi/|\omega/v \mp k_z(\theta)|$ is called the correlation length or the radiation formation zone, which becomes highly oscillating for a fixed but large $z \gg z^{\pm}(\theta)$, once we try to integrate over a range of angles and frequencies, thus its contribution to the total transmitted energy vanishes.

Thus, the observation point's distance from the right- or leftmost slab along z-axis should satisfy the relation:

\begin{gather}
    \frac{\pi}{|\omega/v \mp k_z^{\pm}|}\ll |z-z_{\pm}| \ll \frac{1}{2\Im[k_z^{\pm}]}
\end{gather}

which can be satisfied only if

\begin{gather}
    \varepsilon^{\prime \prime}(\omega) \ll \sqrt{\varepsilon^{\prime}_{\pm}} \cos{\theta_{\pm}}\cdot \bigg|\beta^{-1} \mp \sqrt{\varepsilon^{\prime}_{\pm}} \cos{\theta_{\pm}}\bigg|.
\end{gather}

Apparently, for both forward and backward propagating waves this can be violated at the observation almost perpendicular to the z-axis, but for the forward propagating waves this can be so also near the Cherenkov angle.

\subsubsection{1.1.4 Observables at the Brewster's angle}

Once the condition $\varepsilon^{-1}k_z^{\varepsilon} = b^{-1}k_z^b = \omega/(c\sqrt{\varepsilon+b})$ is satisfied (by the very definition of the Brewster's angle), the intensities, observed at $\theta_{\pm} = \theta_{Br}$, could be drastically simplified: Since

\begin{widetext}
\begin{gather}
    \hat{m}_k = \left( \begin{array}{cc}
        exp(-i(k_z^bd_{2k+1}+k_z^{\varepsilon}d_{2k+2})) & 0 \\
        0 & exp(i(k_z^bd_{2k+1}+k_z^{\varepsilon}d_{2k+2}))
    \end{array} \right), \nonumber\\
    \hat{B} = \frac{2\pi i eq}{cb^{-1}k_z^b}e^{iz_0\omega/v}\left[ \begin{array}{c}
        \left( \frac{b^{-1}}{k_z^b-\omega/v} - \frac{\varepsilon^{-1}}{k_z^{\varepsilon}-\omega/v} \right) \sum_{j=0}^{N_s-1} e^{i\sum_{k=0}^{j-1}[(\omega/v-k_z^b)d_{2k+1}+(\omega/v-k_z^{\varepsilon})d_{2k+2}]} [1-e^{i(\omega/v-k_z^b)d_{2j+1}}] \\
        \left( \frac{b^{-1}}{k_z^b+\omega/v} - \frac{\varepsilon^{-1}}{k_z^{\varepsilon}+\omega/v} \right) \sum_{j=0}^{N_s-1} e^{i\sum_{k=0}^{j-1}[(\omega/v+k_z^b)d_{2k+1}+(\omega/v+k_z^{\varepsilon})d_{2k+2}]} [1-e^{i(\omega/v+k_z^b)d_{2j+1}}]
    \end{array} \right]
    \label{Bmatrix}
\end{gather}
\end{widetext}

we get the amplitudes as $A^+ = -exp(i\sum_{k=0}^{N_s-1}(k_z^bd_{2k+1}+k_z^{\varepsilon}d_{2k+2})) B_1$ and $A^- = B_2$ 

Apparently, the maximum for the intensity is achieved in case of Brewster randomness, when the thickness of each slab is an odd multiplier of the respective correlation length:

\begin{gather}
    (\omega/v\mp k_z^{\varepsilon})d_{2j+2} = \pi(2m_j^{\varepsilon}+1), \nonumber\\ 
    (\omega/v\mp k_z^b)d_{2j+1} = \pi(2m_j^b+1).
\end{gather}

The spectral-angular intensity becomes $\sim N_s^2$.

\subsection{1.2 Periodic Structure}

In case, when these stacks are configured in a periodic structure, so that $d_{2j+1} = a_b$ and $d_{2j+2} = a_{\varepsilon}$ for $\forall j = \overline{0,N_s-1}$ (we can truncate the infinite rightmost medium to a slab with thickness $a_{\varepsilon}$, where only the forward propagating mode survives), with a period $l=a_b+a_{\varepsilon}$, we can write from (\ref{matrixEq}):

\begin{gather}
    \hat{C}_j = e^{ijl\omega/v} \hat{C}_0, \forall j = \overline{0,N_s-1} \nonumber\\ \hat{C}_0 \equiv \hat{C} = \left( \begin{array}{c}
        C_1 \\
        C_2
    \end{array} \right), \nonumber\\ \hat{m}_k = \hat{m} =\left( \begin{array}{cc}
        m_{11} & m_{12} \\
        m_{12}^* & m_{11}^*
    \end{array} \right), \forall k = \overline{0,N_s-1},
\end{gather}

thus 

\begin{gather}
    \hat{B} = \sum_{j=0}^{N_s-1} \hat{m}^j \hat{C}_j,\nonumber\\ \hat{M} = \hat{m}^{N_s}.
\end{gather}

To calculate the powers of matrix $\hat{m}$, the easiest way is to check whether it can be diagonalized. To do this, solve the eigenvalue problem $\hat{m}\cdot \overline{u} = \lambda\cdot \overline{u}$ or:

\begin{gather}
    \left( \begin{array}{cc}
        m_{11}-\lambda & m_{12} \\
        m_{12}^* & m_{11}^* - \lambda
    \end{array} \right) \cdot \left( \begin{array}{c}
        u_1 \\
        u_2
    \end{array} \right) = \left( \begin{array}{c}
        0 \\
        0
    \end{array} \right)
\end{gather}

which can be easily solved:

\begin{gather}
    \lambda = \Gamma \pm \sqrt{\Gamma^2-1}, \nonumber\\ \left( \begin{array}{c}
        u_1 \\
        u_2
    \end{array} \right) = \left( \begin{array}{c}
        m_{12} \\
        \lambda - m_{11}
    \end{array} \right),
\end{gather}

where we used $\det \hat{m} = 1$ and $\Gamma = \Re[m_{11}] = Tr[\hat{m}]/2$. Denote both eigenvalues and their respective eigenvectors by $\lambda_{\pm}$ and $\overline{u}_{\pm}$ and consider the matrix:

\begin{gather}
    \hat{P} = \left( \overline{u}_+,\overline{u}_- \right) = \left( \begin{array}{cc}
        m_{12} & m_{12} \\
        \lambda_+-m_{11} & \lambda_--m_{11}
    \end{array} \right),\nonumber\\
    \hat{m}\cdot \hat{P} = \hat{P}\cdot \left( \begin{array}{cc}
        \lambda_+ & 0 \\
        0 & \lambda_-
    \end{array} \right), \nonumber\\
    \det\hat{P} = m_{12}\cdot \left( \lambda_--\lambda_+ \right) \sim m_{12}\sqrt{\Gamma^2-1}
\end{gather}

The matrix can be inverted if and only if it is not degenerate, which is achieved for $\hat{P}$ if neither $m_{12}=0$ nor $\Gamma = \pm 1$. In this case we have:

\begin{gather}
    \hat{m} = \hat{P}\cdot \left( \begin{array}{cc}
        \lambda_+ & 0 \\
        0 & \lambda_-
    \end{array} \right)\cdot \hat{P}^{-1},\nonumber\\
    \hat{m}^j = \hat{P}\cdot \left( \begin{array}{cc}
        \lambda_+^j & 0 \\
        0 & \lambda_-^j
    \end{array} \right)\cdot \hat{P}^{-1},
\end{gather}

which gives a compact expression for 

\begin{gather}
    \hat{B} = \hat{P} \cdot \left( \begin{array}{cc}
        f_{N_s}(\lambda_+) & 0 \\
        0 & f_{N_s}(\lambda_-)
    \end{array} \right) \cdot \hat{D},\nonumber\\
    \hat{M} = \hat{P}\cdot \left( \begin{array}{cc}
        \lambda_+^{N_s} & 0 \\
        0 & \lambda_-^{N_s}
    \end{array} \right)\cdot \hat{P}^{-1},\nonumber\\
    \hat{D} = \hat{P}^{-1} \hat{C},\nonumber\\
    f_n(\lambda) = \sum_{j=0}^{n-1} \left[ \lambda e^{il\omega/v} \right]^j=\frac{1-\lambda^n exp(inl\omega/v)}{1-\lambda exp(il\omega/v)}.
\end{gather}

and for the amplitudes

\begin{gather}
    A^+ = -\frac{P_{1\alpha}f_{N_s}(\lambda_{\alpha})D_{\alpha}}{P_{1\beta}\lambda^{N_s}_{\beta}(P^{-1})_{\beta 1}},\nonumber\\
    A^- = \left[ P_{2\alpha} - \frac{P_{2\beta}\lambda^{N_s}_{\beta}(P^{-1})_{\beta 1}}{P_{1\gamma}\lambda^{N_s}_{\gamma}(P^{-1})_{\gamma 1}} P_{1\alpha} \right] f_{N_s}(\lambda_{\alpha})D_{\alpha}.
\end{gather}

Here we will focus on their behavior for $N_s \gg 1$. Since $\lambda_+\lambda_-=1$ we have only 2 cases to consider:

\subsubsection{1.2.1 Eigenvalues have different absolute values}

That means that there is an index $M$ so that $|\lambda_M| > 1$, while the other one $\lambda_m = \lambda_M^{-1}$. Emerges when $|\Gamma|>1$.

Expanding in powers of $\lambda_M$, it is easy to get

\begin{gather}
    A^+\approx \frac{exp(iN_sl\omega/v)}{1-\lambda_M exp(il\omega/v)} \frac{D_M}{(P^{-1})_{M1}} + O(\lambda_M^{-N_s}), \nonumber\\ 
    A^-\approx \left( P_{2m} - \frac{P_{2M}}{P_{1M}}P_{1m} \right) \frac{D_m}{1-\lambda_M^{-1}exp(il\omega/v)} +O(\lambda_M^{-N_s})
\end{gather}

\subsubsection{1.2.2 Eigenvalues have the same absolute values}

Denote $\lambda_{\pm} = exp(\mp iq_zl)$, where $q_z$ is determined using equation $\cos(q_zl) = \Gamma$. It represents the quasi-momentum of the Bloch eigenmodes of the electromagnetic waves inside the photonic crystal with the respective periodic structure.

Taking into account, that for $N_s \gg 1$ we have

\begin{gather}
    f_{N_s}(\lambda_{\pm}) = e^{i(N_s-1)l(\omega/v\mp q_z)/2} \frac{\sin{\left[ (\omega/v\mp q_z)lN_s/2\right]}}{\sin{\left[ (\omega/v\mp q_z)l/2\right]}} \approx \nonumber\\ \approx (2\pi/l) \cdot \delta\left( (\omega/v\mp q_z) \mod 2\pi/l \right),
\end{gather}

we get

\begin{widetext}
\begin{gather}
    A^+ \approx -\frac{2\pi}{l} \left[ \frac{P_{11}D_1}{M_{11}} \delta\left( (\omega/v - q_z) \mod 2\pi/l \right) + \frac{P_{12}D_2}{M_{11}} \delta\left( (\omega/v + q_z) \mod 2\pi/l \right) \right],\nonumber\\
    A^- \approx -\frac{2\pi}{l} \left[ \left(P_{21}-\frac{M_{21}}{M_{11}} P_{11}\right)D_1 \delta\left( (\omega/v - q_z) \mod 2\pi/l \right) + \left(P_{22}-\frac{M_{21}}{M_{11}} P_{12}\right)D_2 \delta\left( (\omega/v + q_z) \mod 2\pi/l \right) \right]
\end{gather}
\end{widetext}

which explains sharp peaks near the crossings with the iso-frequency curve, observed in \cite{Lin}.

Now, consider the case $\det\hat{P} = 0$.

\subsubsection{1.2.3 The transfer matrix is already diagonal}

Which means $m_{12} = 0$. This case emerges, when either 1) there is an observation along the Brewster's angle, i. e. $\varepsilon^{-1}k_z^{\varepsilon} = b^{-1}k_z^b$, or 2) for the given observation angle the slab thickness satisfies $\sin{(k_z^ba_b)}=0$.

Applying steps similar to the previous case, we get 

\begin{gather}
    A^+\approx -\frac{2\pi}{l} e^{iq_zlN_s} C_1 \delta\left( (\omega/v - q_z) \mod 2\pi/l \right), \nonumber\\
    A^-\approx \frac{2\pi}{l} C_2 \delta\left( (\omega/v + q_z) \mod 2\pi/l \right).
    \label{mdiagonal}
\end{gather}

\subsubsection{1.2.4 Non-diagonalizable transfer matrix}

Finally, the case $\Gamma = \Re[m_{11}] = \pm 1 \equiv (-1)^A$. In this case, using $\det\hat{m} = |m_{11}|^2 - |m_{12}|^2 = 1$ and $m_{11}=(-1)^A + i\Im[m_{11}]$, it is easy to show that the following matrix

\begin{gather}
    \hat{N} \equiv \hat{m}-(-1)^A {\bf 1} = \left( \begin{array}{cc}
        i(-1)^B|m_{12}| & m_{12} \\
        m_{12}^* & -i(-1)^B|m_{12}|
    \end{array} \right),\nonumber\\
    \hat{N}^2 = \hat{0}
\end{gather}

is nilpotent. This fact allows us to find the powers of $\hat{m}$ as follows:

\begin{gather}
    \hat{m}^j = [(-1)^A {\bf 1} + \hat{N}]^j = (-1)^{Aj}{\bf 1} + j(-1)^{A(j-1)}\hat{N}
\end{gather}

It remains to denote $q \equiv exp(i\alpha) \equiv (-1)^Aexp(il\omega/v)$ to get 

\begin{gather}
    \hat{B} = \left[ \frac{1-q^{N_s}}{1-q} {\bf 1} + (-1)^A \frac{q}{1-q}\left( \frac{1-q^{N_s}}{1-q} - N_s q^{N_s-1} \right) \hat{N} \right]\hat{C}, \nonumber\\
    \hat{M} = (-1)^{AN_s} [{\bf 1} + N_s (-1)^A \hat{N}].
\end{gather}

It is easy to show that even for $N_s|m_{12}| \gg 1$ the amplitudes $A^{\pm}$ cannot become of the order $\sim N_s^2$.

Thus, in all cases, except for the first, $|A^{\pm}| \sim N_s$ and hence the spectral-angular intensity $I \sim N_s^2$.

\subsubsection{1.2.5 Fields inside the PhC}

Now consider the infinite PhC. Decompose the magnetic field as $H_{\Phi} = H_{\Phi}^+ + H_{\Phi}^-$, where

\begin{gather}
    H_{\Phi}^{\pm} (\omega, {\bf q}, z) = \sum_{i=-\infty}^{\infty} [\Theta(z-z_{i-1})-\Theta(z-z_i)] A_i^{\pm} e^{\pm i k_{i,z}(z-z_i)} = \nonumber\\ = \sum_{i=-\infty}^{\infty} [\Theta(z-z_{2i-1})-\Theta(z-z_{2i})] A_{2i}^{\pm} e^{\pm i k_{z}^{\varepsilon}(z-z_{2i})} + \nonumber\\ + \sum_{i=-\infty}^{\infty} [\Theta(z-z_{2i})-\Theta(z-z_{2i+1})] A_{2i+1}^{\pm} e^{\pm i k_{z}^b(z-z_{2i+1})},
\end{gather}

where $A_i^{\pm}$ are connected to those in the neighboring slab through Eq.(\ref{recurrent}), with the absence of any currents, $\hat{B}_i = \hat{0}$. Since 

\begin{gather}
    \left(\begin{array}{c}
        A_{2i}^+ \\
        A_{2i}^-
    \end{array}\right) = \hat{M}_{2i}\hat{M}_{2i+1} \left(\begin{array}{c}
        A_{2(i+1)}^+ \\
        A_{2(i+1)}^-
    \end{array}\right), \nonumber\\
    \left(\begin{array}{c}
        A_{2i-1}^+ \\
        A_{2i-1}^-
    \end{array}\right) = \hat{M}_{2i-1}\hat{M}_{2i} \left(\begin{array}{c}
        A_{2(i+1)-1}^+ \\
        A_{2(i+1)-1}^-
    \end{array}\right), \nonumber\\
    \hat{M}_{2i-1}\hat{M}_{2i} = \hat{M}_{2i+1}\hat{M}_{2i},
\end{gather}

where we used the periodicity of the matrices $\hat{M}_{j+2} = \hat{M}_{j}$, we can derive the following rule:

\begin{widetext}
\begin{gather}
    \left( \begin{array}{c}
        H_{\Phi}^+(z+l) \\
        H_{\Phi}^-(z+l)
    \end{array} \right) = {\bf \hat{M}}(z)\cdot \left( \begin{array}{c}
        H_{\Phi}^+(z) \\
        H_{\Phi}^-(z)
    \end{array} \right), \nonumber\\
    {\bf \hat{M}}(z) = \sum_{i=-\infty}^{\infty} \left( [\Theta(z-z_{2i-1})-\Theta(z-z_{2i})] \times \right. \nonumber\\ \left. \times \left[ \begin{array}{cc}
        \exp(ik_z^{\varepsilon}(z-z_{2i})) & 0 \\
        0 & \exp(-ik_z^{\varepsilon}(z-z_{2i}))
    \end{array} \right]  [\hat{M}_{2i} \hat{M}_{2i+1}]^{-1} \left[ \begin{array}{cc}
        \exp(-ik_z^{\varepsilon}(z-z_{2i})) & 0 \\
        0 & \exp(ik_z^{\varepsilon}(z-z_{2i}))
    \end{array} \right] + \right. \nonumber\\ \left. + [\Theta(z-z_{2i})-\Theta(z-z_{2i+1})] \times \right. \nonumber\\ \left. \times \left[ \begin{array}{cc}
        \exp(ik_z^{b}(z-z_{2i+1})) & 0 \\
        0 & \exp(-ik_z^{b}(z-z_{2i+1}))
    \end{array} \right]  [\hat{M}_{2i+1} \hat{M}_{2i}]^{-1} \left[ \begin{array}{cc}
        \exp(-ik_z^{b}(z-z_{2i+1})) & 0 \\
        0 & \exp(ik_z^{b}(z-z_{2i+1}))
    \end{array} \right] \right),
\end{gather}
\end{widetext}

where $l = a_{\varepsilon}+a_b$. Since $\Gamma = 0.5 \cdot Tr[\hat{M}_{2i} \hat{M}_{2i+1}] = 0.5 \cdot Tr[\hat{M}_{2i+1} \hat{M}_{2i}]$ and $\det[\hat{M}_{2i} \hat{M}_{2i+1}] = \det[\hat{M}_{2i+1} \hat{M}_{2i}] = 1$, the eigenvalues of ${\bf \hat{M}}(z)$ match with those for $\hat{m}^{-1}$. If $|\Gamma| \neq 1$, the matrices are diagonalizable and if $\hat{m} \rightarrow diag(\lambda_+,\lambda_-)$, the matrix ${\bf \hat{M}} \rightarrow diag(\lambda_-,\lambda_+)$. This introduces the forbidden region, determined by $|\lambda_{\pm}| \neq 1$ or $|\Gamma| > 1$, and inside the allowed region it is natural to determine the quasi-momentum $q_z$ through

\begin{gather}
    \lambda_{-} = \Gamma +i\sqrt{1-\Gamma^2} = \exp(+iq_z l), \nonumber\\ q_z l = \arccos\Gamma.
\end{gather}

To find out in which direction this mode is propagating, we need to consider its group velocity \cite{Sakoda}. The latter is always perpendicular to the iso-frequency surface $\omega(q,q_z) = constant$ and in the transverse plane it is directed along the positive direction of $\hat{{\bf q}}$.

\subsection{1.3 Accounting the losses}

If we take into account the possibility of $\Im(\varepsilon) = \varepsilon^{\prime \prime},\Im(b) = b^{\prime \prime} > 0$, the transfer matrix takes a more complicated form:

\begin{widetext}
\begin{gather}
    \hat{m}=\left( \begin{array}{cc}
        e^{-ik_z^{\varepsilon}a_{\varepsilon}}\left[\cos(k_z^ba_b)-\frac{i}{2}\left( \frac{b^{-1}k_z^b}{\varepsilon^{-1} k_z^{\varepsilon}} + \frac{\varepsilon^{-1} k_z^{\varepsilon}}{b^{-1}k_z^b} \right) \sin(k_z^ba_b)\right] & e^{+ik_z^{\varepsilon}a_{\varepsilon}} \left(-\frac{i}{2}\right)\left( \frac{b^{-1}k_z^b}{\varepsilon^{-1} k_z^{\varepsilon}} - \frac{\varepsilon^{-1} k_z^{\varepsilon}}{b^{-1}k_z^b} \right) \sin(k_z^ba_b) \\
        e^{-ik_z^{\varepsilon}a_{\varepsilon}} \left(+\frac{i}{2}\right)\left( \frac{b^{-1}k_z^b}{\varepsilon^{-1} k_z^{\varepsilon}} - \frac{\varepsilon^{-1} k_z^{\varepsilon}}{b^{-1}k_z^b} \right) \sin(k_z^ba_b) & e^{+ik_z^{\varepsilon}a_{\varepsilon}}\left[\cos(k_z^ba_b)+\frac{i}{2}\left( \frac{b^{-1}k_z^b}{\varepsilon^{-1} k_z^{\varepsilon}} + \frac{\varepsilon^{-1} k_z^{\varepsilon}}{b^{-1}k_z^b} \right) \sin(k_z^ba_b)\right]
    \end{array} \right)
\end{gather}
\end{widetext}

However, its eigenvalues are still determined by $\lambda = \Gamma \pm\sqrt{\Gamma^2-1}, \text{ } \Gamma = (1/2) Tr[\hat{m}]$. Here we will take into account only the small losses which nonetheless can impose a restriction on total number of slabs in a stack.

Keeping only the corrections of the first order, we get $\Gamma = \Gamma^{\prime} + i\Gamma^{\prime \prime}$:

\begin{widetext}
\begin{gather}
    \Gamma^{\prime} \approx \cos(k_z^{\varepsilon \prime}a_{\varepsilon}) \cos(k_z^{b \prime} a_b) - \frac{1}{2}\left( \frac{b^{\prime -1}k_z^{b \prime}}{\varepsilon^{-1 \prime} k_z^{\varepsilon \prime}} + \frac{\varepsilon^{\prime -1} k_z^{\varepsilon \prime}}{b^{\prime -1}k_z^{b \prime}} \right) \sin(k_z^{\varepsilon \prime}a_{\varepsilon}) \sin(k_z^{b \prime} a_b), \nonumber\\ 
    \Gamma^{\prime \prime} \approx \varepsilon^{\prime \prime} \left[ \frac{1}{2}\left( \frac{b^{\prime -1}k_z^{b \prime}}{\varepsilon^{-1 \prime} k_z^{\varepsilon \prime}} - \frac{\varepsilon^{\prime -1} k_z^{\varepsilon \prime}}{b^{\prime -1}k_z^{b \prime}} \right) \sin(k_z^{\varepsilon \prime}a_{\varepsilon}) \sin(k_z^{b \prime} a_b) \left( \varepsilon^{\prime -1} - \frac{1}{2\sqrt{\varepsilon^{\prime} -(cq/\omega)^2}}\right) - \right.\nonumber\\ \left. - \frac{\omega a_{\varepsilon}}{2c\sqrt{\varepsilon^{\prime} -(cq/\omega)^2}} \left( \sin(k_z^{\varepsilon \prime}a_{\varepsilon}) \cos(k_z^{b \prime} a_b) - \frac{1}{2}\left( \frac{b^{\prime -1}k_z^{b \prime}}{\varepsilon^{-1 \prime} k_z^{\varepsilon \prime}} + \frac{\varepsilon^{\prime -1} k_z^{\varepsilon \prime}}{b^{\prime -1}k_z^{b \prime}} \right) \cos(k_z^{\varepsilon \prime}a_{\varepsilon}) \sin(k_z^{b \prime} a_b) \right) \right] + \nonumber\\ + b^{\prime \prime} \left[ \frac{1}{2}\left( \frac{b^{\prime -1}k_z^{b \prime}}{\varepsilon^{-1 \prime} k_z^{\varepsilon \prime}} - \frac{\varepsilon^{\prime -1} k_z^{\varepsilon \prime}}{b^{\prime -1}k_z^{b \prime}} \right) \sin(k_z^{\varepsilon \prime}a_{\varepsilon}) \sin(k_z^{b \prime} a_b) \left( \frac{1}{2\sqrt{b^{\prime} -(cq/\omega)^2}} - b^{\prime -1} \right) - \right.\nonumber\\ \left. - \frac{\omega a_b}{2c\sqrt{b^{\prime} -(cq/\omega)^2}} \left( \cos(k_z^{\varepsilon \prime}a_{\varepsilon}) \sin(k_z^{b \prime} a_b) - \frac{1}{2}\left( \frac{b^{\prime -1}k_z^{b \prime}}{\varepsilon^{-1 \prime} k_z^{\varepsilon \prime}} + \frac{\varepsilon^{\prime -1} k_z^{\varepsilon \prime}}{b^{\prime -1}k_z^{b \prime}} \right) \sin(k_z^{\varepsilon \prime}a_{\varepsilon}) \cos(k_z^{b \prime} a_b) \right) \right].
\end{gather}
\end{widetext}

Since we are primarily interested in the region $|\Gamma^{\prime}| < 1$, we get the following for the eigenvalues:

\begin{eqnarray}
    \lambda \approx \left(\Gamma^{\prime} \pm i \sqrt{1-\Gamma^{\prime 2}}\right) \left[ 1 \pm \frac{\Gamma^{\prime \prime}}{\sqrt{1-\Gamma^{\prime 2}}} \right]
\end{eqnarray}

which suggests that $f_{N_s}(\lambda) \approx N_s$ can be achieved only if 

\begin{gather}
    N_s \ll \frac{\sqrt{1-\Gamma^{\prime 2}}}{|\Gamma^{\prime \prime}|}
\end{gather}

Thus, absorption puts certain restrictions on the maximum number of slabs. It can be proved that in the case $|\Gamma^{\prime}| = 1$ that number is restricted by $|\Gamma^{\prime \prime}|^{-1/2}$, which is much less than in the previous case.

\subsection{1.4 More on observations along the Brewster's angle}

Assume our photonic crystal consists of materials that are non-dispersive in the frequency region of interest. With this

\begin{gather}
    \hat{C} = \frac{2\pi i e}{\omega} e^{iz_0 \omega/v} {\bf \hat{C}}(\tilde{a}_b,\tilde{a}_{\varepsilon},b,\varepsilon,\theta), \nonumber\\
    \hat{B} = \sum_{j=0}^{N_s-1} \left(e^{il\omega/v} \hat{m} \right)^j \hat{C} = \frac{2\pi i e}{\omega} e^{iz_0 \omega/v} {\bf \hat{B}}(\tilde{a}_b,\tilde{a}_{\varepsilon},b,\varepsilon,\theta),
\end{gather}

where

\begin{widetext}
\begin{gather}
    {\bf \hat{C}} = \sqrt{\varepsilon^{\prime}} \sin \theta \left[ (1-\exp(i 2\pi \tilde{a}_b \beta^{-1})\cos(2\pi \tilde{a}_b \sqrt{b - \varepsilon^{\prime} \sin^2 \theta})) \times \right. \nonumber\\ \left. \times \left( \frac{1}{b-\varepsilon^{\prime}\sin^2 \theta -\beta^{-2}} \left[ \begin{array}{c}
        1+\frac{\beta^{-1} b^{-1}}{\varepsilon^{-1} \sqrt{\varepsilon-\varepsilon^{\prime}\sin^2 \theta}} \\
        1-\frac{\beta^{-1} b^{-1}}{\varepsilon^{-1} \sqrt{\varepsilon-\varepsilon^{\prime}\sin^2 \theta}}
    \end{array} \right] - \frac{1}{\varepsilon-\varepsilon^{\prime}\sin^2 \theta -\beta^{-2}} \left[ \begin{array}{c}
        1+\frac{\beta^{-1} }{ \sqrt{\varepsilon-\varepsilon^{\prime}\sin^2 \theta}} \\
        1-\frac{\beta^{-1} }{ \sqrt{\varepsilon-\varepsilon^{\prime}\sin^2 \theta}}
    \end{array} \right] \right) + \right. \nonumber\\ \left. + i\exp (i 2\pi \tilde{a}_b \beta^{-1}) \sin(2\pi \tilde{a}_b \sqrt{b - \varepsilon^{\prime} \sin^2 \theta}) \times \right. \nonumber\\ \left. \times \left( \frac{1}{b-\varepsilon^{\prime}\sin^2 \theta -\beta^{-2}} \left[ \begin{array}{c}
        \frac{ \beta^{-1} }{ \sqrt{b-\varepsilon^{\prime}\sin^2 \theta} }+\frac{ b^{-1} \sqrt{b-\varepsilon^{\prime}\sin^2 \theta} }{\varepsilon^{-1} \sqrt{\varepsilon-\varepsilon^{\prime}\sin^2 \theta}} \\
        \frac{ \beta^{-1} }{ \sqrt{b-\varepsilon^{\prime}\sin^2 \theta} }-\frac{ b^{-1} \sqrt{b-\varepsilon^{\prime}\sin^2 \theta} }{\varepsilon^{-1} \sqrt{\varepsilon-\varepsilon^{\prime}\sin^2 \theta}}
    \end{array} \right] - \frac{1}{\varepsilon-\varepsilon^{\prime}\sin^2 \theta -\beta^{-2}} \left[ \begin{array}{c}
        \frac{ \beta^{-1} \varepsilon^{-1} }{ b^{-1} \sqrt{b-\varepsilon^{\prime}\sin^2 \theta} }+\frac{ b^{-1} \sqrt{b-\varepsilon^{\prime}\sin^2 \theta} }{\varepsilon^{-1} \sqrt{\varepsilon-\varepsilon^{\prime}\sin^2 \theta}} \\
        \frac{ \beta^{-1} \varepsilon^{-1} }{ b^{-1} \sqrt{b-\varepsilon^{\prime}\sin^2 \theta} }-\frac{ b^{-1} \sqrt{b-\varepsilon^{\prime}\sin^2 \theta} }{\varepsilon^{-1} \sqrt{\varepsilon-\varepsilon^{\prime}\sin^2 \theta}}
    \end{array} \right] \right) \right], \nonumber\\ {\bf \hat{B}} = \sum_{j=0}^{N_s-1} \left(e^{i2\pi \tilde{l} \beta^{-1}} \hat{m} \right)^j {\bf \hat{C}}.
\end{gather}
\end{widetext}

and we used $l = a_b + a_{\varepsilon}$ and $\tilde{a}_X = a_X/ \lambda, \text{ } X = b,\varepsilon$. This leads to 

\begin{gather}
    |A^{\pm}|^2 = \frac{4\pi^2 e^2}{\omega^2} |\alpha^{\pm}(\tilde{a}_b,\tilde{a}_{\varepsilon},b,\varepsilon,\theta)|^2, \nonumber\\ \frac{1}{\hbar} I^{r,\pm} (\omega, \theta) = \alpha \frac{\sqrt{ \varepsilon^{\prime} } \cos^2 \theta }{4\pi^2} |\alpha^{\pm}(\tilde{a}_b,\tilde{a}_{\varepsilon},b,\varepsilon,\theta)|^2
\end{gather}

where $\alpha = e^2/\hbar c \approx 1/ 137$ is the fine-structure constant.

The following results lead to specific scaling rules for the angular intensity ${\bf I}(\theta) = \int d\omega I(\omega,\theta)$. If we could extrapolate the integration over all possible (positive) frequencies, we would get for the crystal, scaled by the factor $x>0$: $(a_b, a_{\varepsilon}) \rightarrow (xa_b, xa_{\varepsilon})$, we get for the angular intensity:

\begin{gather}
    {\bf I}(\theta;xa_b, xa_{\varepsilon}) = \int_0^{\infty} d\omega I(\omega x a_b, \omega xa_{\varepsilon}) = \nonumber\\ \frac{1}{x} \int_0^{\infty} d\omega^{\prime} I(\omega^{\prime} a_b, \omega^{\prime} a_{\varepsilon}) = \frac{1}{x} {\bf I}(\theta;a_b,a_{\varepsilon}),
\end{gather}

where we performed the linear transformation of the integration variable $\omega \rightarrow \omega^{\prime} = x \omega$. This suggests implementation of PhC with possibly the smallest period $l$. 

For $N_s \gg 1$ highly oscillating terms can emerge solely from the part $\sum_{j=0}^{N_s-1} \left(e^{i2\pi \tilde{l} \beta^{-1}} \hat{m} \right)^j$. If we exclude the special case $|\Gamma| \neq 1$, the matrix $\hat{m}$ becomes diagonalizable and the term mentioned above reduces to a combination of $f_{N_s}(\lambda_{\pm})$. Thus, we proceed with considering the oscillatory behavior of these terms. We will consider the case of absence of any absorption, since in a more general case of a small absorption the main results could be retrieved by replacement $\varepsilon \rightarrow \varepsilon^{\prime}$ and $b \rightarrow b^{\prime}$.

Consider the approximations for ultra-relativistic particles ($\gamma \gg 1$). In this case, along with $ \tilde{q}_z = q_z/(\omega/c) = \tilde{q}_z(\tilde{a}_b,\tilde{a}_{\varepsilon},b,\varepsilon,\theta)$ and $\tilde{l} = l/\lambda$, we get

\begin{eqnarray}
    f_{N_s}(\lambda_+ = \exp(-q_zl)) \approx \nonumber\\ \approx \frac{1-\exp(2\pi N_s\tilde{l}(1-\tilde{q}_z + \gamma^{-2}/2))}{1-\exp(2\pi \tilde{l}(1-\tilde{q}_z + \gamma^{-2}/2))},
    \label{forward}
\end{eqnarray}

where we used $ c/v =\beta^{-1} \approx 1+\gamma^{-2}/2$. If PhC allows for the fulfillment of condition $\tilde{q}_z \equiv 1$ for any frequency $\omega = 2\pi c/\lambda$ in the region of interest, the exponent in the denominator will stay close to 1, if $\tilde{l} \ll \gamma^2$ in the region of interest. This condition is automatically fulfilled, since $\tilde{q}_z \in (0,1/2\tilde{l})$ and once $\tilde{q}_z = 1$, it requires $\tilde{l} < 1/2 \ll \gamma^2$.

At Brewster's angle $q_z l = \pm (k_z^ba_b+k_z^{\varepsilon}a_{\varepsilon}) + 2\pi n$, where the sign in front of the brackets and the integer $n$ should be chosen so that $q_zl \in (0,\pi)$. With this

\begin{eqnarray}
    \tilde{q}_z = \pm \frac{ba_b+\varepsilon a_{\varepsilon}}{l\sqrt{b + \varepsilon}} + \frac{n}{\tilde{l}}
\end{eqnarray}

Only the second term in the rhs. is $\lambda$-dependent, unless $n = 0$. The first term can be $=1$ only in case of the $+$ sign (satisfied automatically, since it is required to be $\tilde{q}_z > 0$ by the very definition), and

\begin{eqnarray}
    \frac{a_b}{a_{\varepsilon}}=\frac{\sqrt{\varepsilon+b}-\varepsilon}{b-\sqrt{\varepsilon+b}}.
    \label{ratio}
\end{eqnarray}

The rhs. of this equation can be positive only in two cases:

\begin{gather}
\text{ either } \nonumber
    b > \max\left( \frac{1}{2}+\sqrt{\varepsilon+\frac{1}{4}}, \varepsilon (\varepsilon-1) \right) \nonumber\\ \text{ or } b < \min\left( \frac{1}{2}+\sqrt{\varepsilon+\frac{1}{4}}, \varepsilon (\varepsilon-1) \right)
\end{gather}

\begin{figure*}
    \centering
    \includegraphics[width=0.9\linewidth]{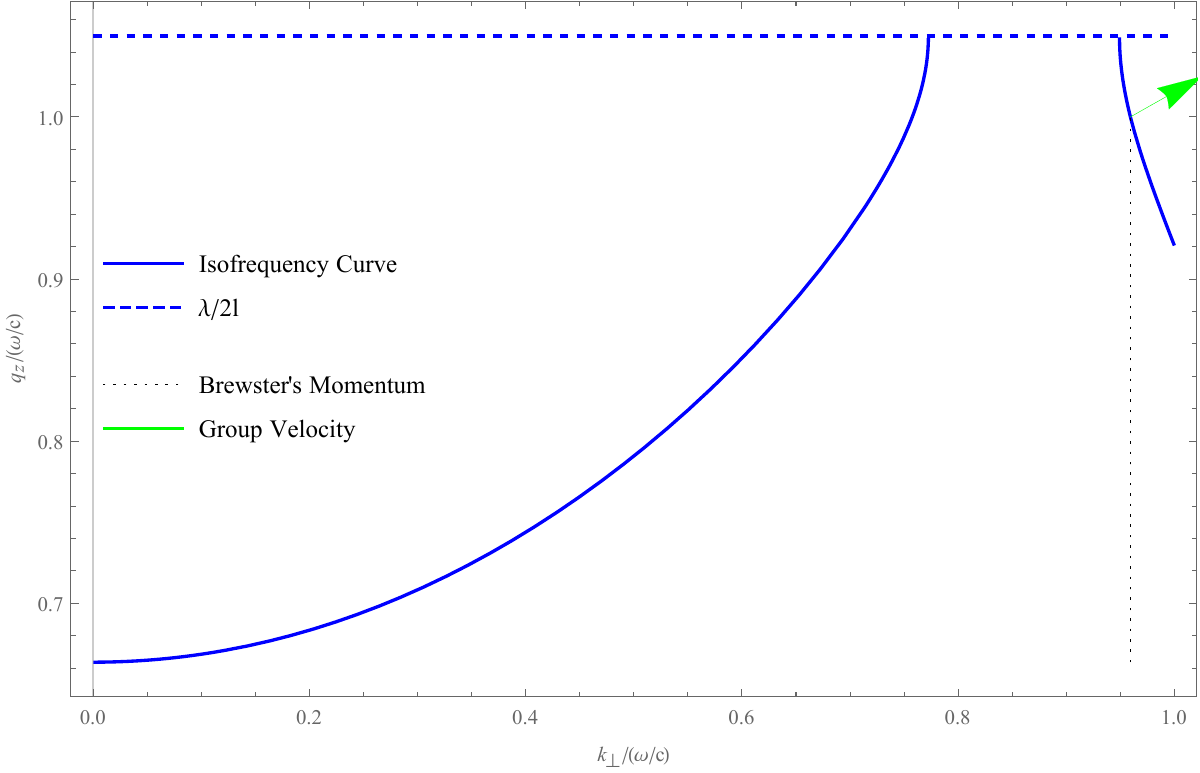}
    \caption{Iso-frequency curve of the photonic crystal with $\lambda = 2.1 l$ and $a_b/a_{\varepsilon}$ given by Eq.(\ref{ratio}) with $b = 3.4$ and $\varepsilon = 1$. The green arrow indicates the direction of $\vec{v}_g$ for given transversal momentum $k_{\perp} = q$.}
    \label{isofreqvgroup}
\end{figure*}

From now on, we consider the case $\varepsilon=1$. Interestingly, this condition can be satisfied for the dielectrics above a threshold dielectric constant, given by $b_{threshold} = (1+\sqrt{5})/2 \approx 1.618...$ the Golden ratio. Also, the numerical simulation in Fig.(\ref{isofreqvgroup}) shows that the z-component of the group velocity of the waves, corresponding to the eigenvalue $exp(-iq_z l)$, is nonnegative once $\tilde{l} < 1/2$. This shows that modes, corresponding to the eigenvalue $\lambda_+$ indeed propagate in positive z-direction.

Through Eq.(\ref{Bmatrix}) for the periodic system with the structure given in (\ref{ratio}) we get the forward-propagating amplitude:

\begin{eqnarray}
    A^+ \sim \frac{1-exp(iN_s\omega l(v^{-1}-c^{-1}))}{1-exp(i\omega l(v^{-1}-c^{-1}))}
\end{eqnarray}

\subsubsection{1.4.1 Saturation region}

For ultrarelativistic particles $\gamma \gg 1$ and $v^{-1}-c^{-1} \approx \gamma^{-2}/(2c)$. We can write $A^+ \sim N_s$ once $N_s\omega l/(2c\gamma^2)\ll 1$ or $\gamma \gg \sqrt{N_s\omega l/2c}$.

\begin{figure*}
    \centering
    \includegraphics[width=0.9\linewidth]{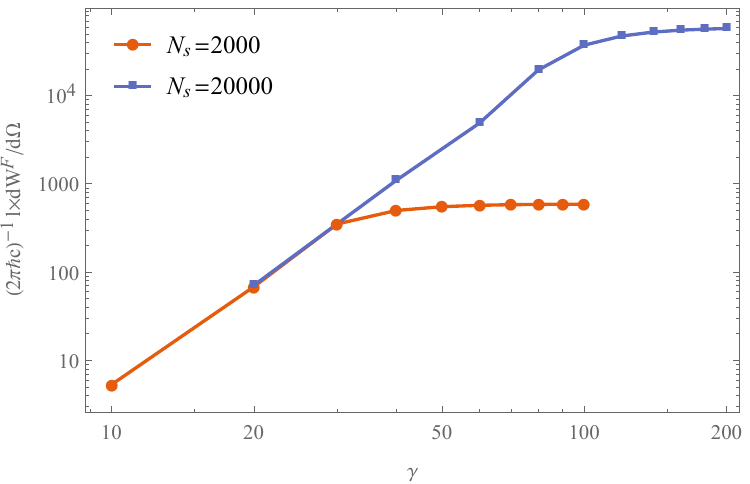}
    \caption{Saturation effect for $N_s = 2000$ and $20000$ respectively. The intensity was obtained after integration over wavelengths from $\lambda/l = 2$ to $\lambda/l = 50$.}
    \label{Saturation}
\end{figure*}

As can be seen, this threshold for the Lorentz factor increases along with $N_s$. Above the threshold there is a saturation effect, where the spectral angular intensity is no more $\gamma$ dependent (see Fig.(\ref{Saturation})).

\subsubsection{1.4.2 Below the saturation region}

Consider the region below the threshold, where the particle is still ultra-relativistic: $1 \ll \gamma \ll \sqrt{\pi N_s \tilde{l}}$. Then

\begin{gather}
    \pi N_s \tilde{l} \gamma^{-2} \gg 1 \rightarrow \text{ oscillating numerator,} \nonumber \\
    \pi \tilde{l} \gamma^{-2} \ll 1 \rightarrow \text{ stable exponent in denominator}
\end{gather}

in Eq.(\ref{forward}). Integral in the frequency region $\omega \in (2\pi c/\lambda_{\max}, 2\pi c/\lambda_{\min})$, where $\lambda_{\min} > 2l$ and all materials should be non-dispersive in the region between maximal and minimal frequencies.

Since we have

\begin{gather}
    I^{+}(\omega,\theta_{Br}) = \alpha \hbar \frac{\sqrt{\varepsilon}\cos^2\theta_{Br}}{4\pi^2} |f_{N_s}(\exp(-iq_z l))|^2 |[{\bf \hat{C}}(\theta_{Br})]_{1}|, \nonumber\\
    I^{-}(\omega,\theta_{Br}) = \alpha \hbar \frac{\sqrt{\varepsilon}\cos^2\theta_{Br}}{4\pi^2} |f_{N_s}(\exp(+iq_z l))|^2 |[{\bf \hat{C}}(\theta_{Br})]_{2}|,
\end{gather}

and for estimation purpose we can represent the integral over frequencies of a product of highly and slowly oscillating functions $F$ and $C$ respectively as $\int d\omega F(\omega) C(\omega) = C(\omega^*) \int d\omega F(\omega)$, where $\omega^* \in (\omega_{\min}, \omega_{\max})$, it remains for us to consider the properties of the integrals $\int d\omega |f_{N_s}(\exp(\mp q_z l))|^2$. Since $\tilde{l} < 1/2 \rightarrow \omega_{\max} < \pi c/l$ and

\begin{gather}
    \int_{\omega_{\min}}^{\omega_{\max}} d\omega |f_{N_s}(\exp(-iq_z l))|^2 \approx \nonumber\\ \approx 2 \int_{\omega_{\min}}^{\omega_{\max}} d\omega \frac{1-\cos\left(N_s \frac{\omega l}{2c\gamma^2} \right)}{\left( \frac{\omega l}{2c\gamma^2} \right)^2} =\nonumber\\ = \frac{4c\gamma^2}{l} \int_{y_{\min}}^{y_{\max}}dy \frac{1-\cos N_s y}{y^2},
\end{gather}

where $y = \omega l/2c\gamma^2 = \pi \tilde{l} \gamma^{-2}$. Since we have 

\begin{gather}
    \frac{1-\cos N_s y}{y^2} dy = d\left(-\frac{1-\cos N_s y}{y}\right)+N_s\frac{\sin N_s y}{y} dy, \nonumber\\
    \lim_{N_s\rightarrow \infty} \frac{\sin N_s y}{y} = \pi \delta(y),
\end{gather}

and $y_{\min} > 0$, we get finally 

\begin{gather}
    \int_{\omega_{\min}}^{\omega_{\max}} d\omega |f_{N_s}(\exp(-iq_z l))|^2 \approx \frac{4 c}{\pi l} \gamma^4 \left[ \frac{1-\cos( N_s\pi \tilde{l}_{\min} \gamma^{-2})}{\tilde{l}_{\min}} - \right. \nonumber\\ \left. -\frac{1-\cos( N_s\pi \tilde{l}_{\max} \gamma^{-2})}{\tilde{l}_{\max}} \right].
\end{gather}

\begin{figure*}
    \centering
    \includegraphics[width=0.9\linewidth]{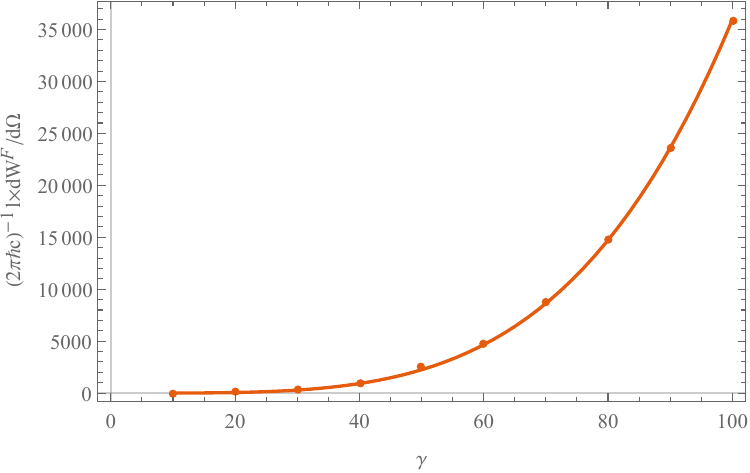}
    \caption{Plot for the forward propagating waves in the region before saturation, $N_s = 2\times 10^6$. The intensity was obtained after integration over wavelengths from $\lambda/l = 2$ to $\lambda/l = 20$.}
    \label{BeforeSatF}
\end{figure*}

Thus, for the waves, propagating in the forward direction at Brewster's angle wrt. z-axis, we have for the angular intensity ${\bf I}^+(\theta = \theta_{Br}) \sim \gamma^4$ (see Fig.(\ref{BeforeSatF})).

\begin{figure*}
    \centering
    \includegraphics[width=0.9\linewidth]{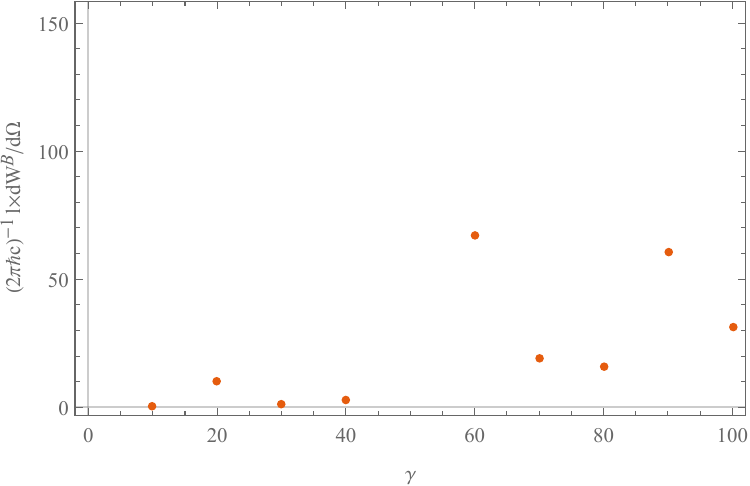}
    \caption{Plot for the backward propagating waves in the region before saturation, $N_s = 2\times 10^6$. The intensity was obtained after integration over wavelengths from $\lambda/l = 2$ to $\lambda/l = 20$.}
    \label{BeforeSatB}
\end{figure*}

Now, for the second integral

\begin{gather}
    \int_{\omega_{\min}}^{\omega_{\max}} d\omega |f_{N_s}(\exp(+iq_z l))|^2 \approx \int_{\omega_{\min}}^{\omega_{\max}} d\omega \Bigg|\frac{1-\exp(iN_s\omega \cdot 2l/c)}{1-\exp(i\omega \cdot 2l/c)}\Bigg|^2 \nonumber\\ = \frac{c}{2l} \int_{y_{\min}}^{y_{\max}}dy \Bigg| \frac{1-\exp(iN_s y)}{1-\exp(iy)} \Bigg|^2 \rightarrow\frac{c}{2l} N_s \int_{y_{\min}}^{y_{\max}}dy \delta(y \mod 2\pi)
\end{gather}

, where $y = \omega \cdot 2l/c$. Taking into account, that $y_{\max} < 2\pi$ and $y_{min}>0$, the integral of $\delta$-function vanishes: ${\bf I^-} (\theta = \theta_{Br})\approx 0$ (see Fig.(\ref{BeforeSatB})).

\subsubsection{1.4.3 Small Angular Deviations}

Now consider the vicinity of the Brewster's angle. We will neglect the corrections of order $O(\delta \theta^2)$, keeping only linear terms. Thus, with this approximation

\begin{gather}
    \delta\left( \frac{\varepsilon^{-1}k_z^{\varepsilon}}{b^{-1}k_z^b} +\frac{b^{-1}k_z^b}{\varepsilon^{-1}k_z^{\varepsilon}} \right) \approx 0,
\end{gather}

since near the Brewster's angle $\varepsilon^{-1}k_z^{\varepsilon}=b^{-1}k_z^b$. Hence $\delta \Gamma = \delta [\cos (k_z^{\varepsilon}a_{\varepsilon} + k_z^b a_b)]$ and 

\begin{gather}
    \delta \tilde{q}_z = -\delta\theta \sqrt{\frac{\varepsilon}{b}}[\sqrt{\varepsilon+b}-1] = \tilde{q}_z - 1
\end{gather}

With this the highly oscillatory function

\begin{gather}
    f_{N_s}(\exp(-iq_z l)) \approx \frac{1-\exp(i2\pi\tilde{l}N_s(-\delta \tilde{q}_z + \gamma^{-2}/2))}{1-\exp(i2\pi\tilde{l}(-\delta \tilde{q}_z + \gamma^{-2}/2))}
\end{gather}

achieves its maximum at $\delta \theta_{\gamma} = -\frac{1}{2\gamma^2} \frac{\sqrt{b/\varepsilon}}{\sqrt{\varepsilon+b}-1}$, and its width is independent of $\gamma$: introducing $\delta \theta^{\prime} = \delta\theta + \delta\theta_{\gamma} = \delta\theta - |\delta\theta_{\gamma}|$, we require 
\begin{gather}
    |\delta\theta^{\prime}|\ll \frac{\sqrt{b/\varepsilon}}{\sqrt{\varepsilon+b}-1}.
\end{gather}

for the denominator to be stable, as in previous cases.

\subsubsection{1.4.4 Deviations from the proportion in Eq.(\ref{ratio})}

If we consider deviations from the proportion $f_0^{\varepsilon} = a_{\varepsilon}^0/l = (b - \sqrt{b+\varepsilon})/(b-\varepsilon) \rightarrow f^{\varepsilon} = a_{\varepsilon}/l = f_0^{\varepsilon} - \delta f$, hence $f_b = 1-f_{\varepsilon} = f_b^0 + \delta f$, we will get

\begin{gather}
    \tilde{q}_z = 1 + \frac{b-\varepsilon}{\sqrt{\varepsilon+b}}\delta f.
\end{gather}

This shifts the peak of the spectral-angular intensity to those $\gamma$, which are determined by 

\begin{gather}
    \frac{\gamma^{-2}}{2} - \frac{b-\varepsilon}{\sqrt{\varepsilon+b}} \delta f
= 0\end{gather}

For $\delta f > 0$, it has a real solution, and the peak width is determined by 

\begin{gather}
    N_s \tilde{l} \bigg| \frac{\gamma^{-2}}{2} - \frac{b-\varepsilon}{\sqrt{\varepsilon+b}} \delta f \bigg| \sim 1.
\end{gather}

Outside that region, determined by the relation

\begin{gather}
    N_s \tilde{l} \bigg| \frac{\gamma^{-2}}{2} - \frac{b-\varepsilon}{\sqrt{\varepsilon+b}} \delta f \bigg| \gg 1 \gg \tilde{l} \bigg| \frac{\gamma^{-2}}{2} - \frac{b-\varepsilon}{\sqrt{\varepsilon+b}} \delta f \bigg|
\end{gather}

we have a highly oscillating numerator and a slowly varying denominator in $I(\omega, \theta)$, which, after integration over frequencies, leads to

\begin{gather}
    {\bf I}^+ \sim \frac{\gamma^4}{\bigg| \frac{1}{2} - \frac{b-\varepsilon}{\sqrt{b+\varepsilon}} \delta f\cdot \gamma^2 \bigg|^2}.
\end{gather}

\subsubsection{1.4.5 Dealing with deviations $\delta f$}

Considering the contributions of deviations from $\theta_{Br}$ and the ideal proportion $f_0^{\varepsilon,b}$ to the phase:

\begin{gather}
    \delta \tilde{q}_z = \frac{b-\varepsilon}{\sqrt{\varepsilon+b}}\delta f - \sqrt{\frac{\varepsilon}{b}}[\sqrt{\varepsilon+b}-1] \delta \theta,
\end{gather}

we immediately note, that the linear terms could be cancelled out once we choose 

\begin{gather}
    \delta \theta = \frac{b-\varepsilon}{\sqrt{\varepsilon+b}[\sqrt{\varepsilon+b}-1]} \sqrt{\frac{b}{\varepsilon}} \delta f.
\end{gather}

This means that small deviations from the ideal proportion shift the angular position of the peak according to the formula above. Inside the detector, the influence of $\delta f$ could be cancelled out by slightly shifting the observation direction from the $\theta_{Br}$.

\clearpage 